\documentclass{article}[12pt]
\usepackage{setspace}
\usepackage{colortbl,xcolor} 
\usepackage{arxiv}

\usepackage[utf8]{inputenc} % allow utf-8 input
\usepackage[T1]{fontenc}    % use 8-bit T1 fonts
\usepackage{hyperref}
\usepackage{geometry}
\usepackage{array}
\usepackage{booktabs}
\usepackage{caption}
\usepackage{url}            % simple URL typesetting
\usepackage{booktabs}       % professional-quality tables
\usepackage{amsfonts}       % blackboard math symbols
\usepackage{nicefrac}       % compact symbols for 1/2, etc.
\usepackage{microtype}      % microtypography
\usepackage{lipsum}	
\usepackage{bbm}
\usepackage{xcolor}
\usepackage{amsthm}
\usepackage{pdflscape}
\usepackage{pgfplots}
\usepackage{mathrsfs}
\usepackage{mathtools}
\usepackage{algorithmic}
\usepackage{algorithm}

\theoremstyle{plain}

\newtheorem*{theorem*}{Theorem}

\newtheorem*{proposition}{Proposition}

\theoremstyle{definition}

\theoremstyle{remark}

\newtheorem*{remark}{Remark}

\newcommand{\dottedcolumn}[3]{%
  \settowidth{\dimen0}{$#1$}
  \settowidth{\dimen2}{$#2$}
  \ifdim\dimen2>\dimen0 \dimen0=\dimen2 \fi
  \begin{pmatrix}\,
    \vcenter{
      \kern.6ex
      \vbox to \dimexpr#1\normalbaselineskip-1.2ex{
        \hbox{$#2$}
    \kern3pt
    \xleaders\vbox{\hbox to \dimen0{\hss.\hss}\vskip4pt}\vfill
    \kern1pt
    \hbox{$#3$}
  }\kern.6ex}\,
  \end{pmatrix}
}

\title{Copula-Based Synthetic Data Generation: A Use-Inspired Perspective on Trade-offs}

\author{ \href{https://orcid.org/0000-0000-0000-0000}{\includegraphics[scale=0.05]{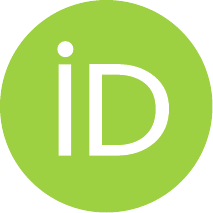}\hspace{1mm}Antoine ~Faul}\thanks{corresponding author} \\
	Institute of Mathematical Statistics\\ and Actuarial Sciences\\
	University of Bern, Switzerland\\
	\texttt{antoine.faul@unibe.ch} \\
	\And
	\href{https://orcid.org/0000-0000-0000-0000}{\includegraphics[scale=0.06]{orcid.pdf}\hspace{1mm}David ~Ginsbourger} \\
	Institute of Mathematical Statistics \\ and Actuarial Sciences\\
	University of Bern, Switzerland\\
	\texttt{david.ginsbourger@unibe.ch} \\
 \And
	\href{https://orcid.org/0000-0000-0000-0000}{\includegraphics[scale=0.05]{orcid.pdf}\hspace{1mm}Ben  ~Spycher} \\
	Institute of Social and Preventive Medicine\\
	University of Bern, Switzerland\\
	\texttt{ben.spycher@unibe.ch} \\
    \And
	\href{https://orcid.org/0000-0000-0000-0000}{\includegraphics[scale=0.05]{orcid.pdf}\hspace{1mm}Petra  ~Stute} \\
	Department of Obstetrics and Gynecology\\
	University Women's Hospital, Bern, Switzerland\\
	\texttt{petra.stute@insel.ch} \\
}

\date{}

\renewcommand{\headeright}{Faul \& al.}
\renewcommand{\shorttitle}{Copula based Synthetic Data Generation}

\hypersetup{
pdftitle={A template for the arxiv style},
pdfsubject={q-bio.NC, q-bio.QM},
pdfauthor={David S.~Hippocampus, Elias D.~Striatum},
pdfkeywords={First keyword, Second keyword, More},
}

\begin{document}
\maketitle

\begin{abstract}

The generation of synthetic data offers a promising way to address challenges related to data scarcity and privacy concerns. While many approaches have been developed, tabular synthetic data generation methods often require complex models and extensive training data, while lacking interpretability. In this work, we investigate versatile probabilistic approaches using copulas — and leverage in particular the Gaussian Mixture Copula Model — to model the joint distribution of multivariate data, with a main focus on moderate-dimensional data sets. Taking a use-inspired perspective, we examine key trade-offs inherent to synthetic data evaluation — spanning \textit{fidelity}, \textit{utility}, and \textit{privacy} — and demonstrate that copula-based methods achieve a sound balance across all three dimensions while remaining interpretable and providing an explicit probabilistic model. We illustrate and highlight the potential benefits of the method on a test case focusing on systematic missing values in a cardiovascular risk calculator and develop specific evaluation metrics for this application.
\end{abstract}

% keywords can be removed
%\noindent%
%{\it Keywords:} Synthetic data generation, Copulas, Evaluation metrics, Privacy

\section{Introduction}
In medical research, clinicians work with valuable patient data that could be instrumental for tackling important research questions. However, privacy concerns often delay or restrict data sharing with other researchers.
Synthetic data generation offers a compelling compromise by creating datasets that mirror the statistical properties of the original data while enhancing privacy. 
Researchers can  then conduct statistical analyses on resulting synthetic data, allowing for developing reproducible models that may eventually be applied to patient data once conditions are met.

The field of synthetic data generation has expanded rapidly in recent years, leading to the development of numerous machine learning methods for tabular data synthesis \cite{xu2019modeling,jordon2018pate,watson2023adversarial}. These models offer significant potential but may often be challenging to interpret and require large training datasets. Furthermore, they are susceptible to overfitting, which can compromise privacy by inadvertently replicating specific patient details instead of capturing general patterns.

While increasing the complexity of generators usually enhances synthetic data \textit{fidelity} - how closely the synthetic data statistically resembles
the real data - and \textit{utility} - the usefulness of synthetic data for a downstream application - it tends to deteriorate \textit{privacy} - how much information the synthetic data discloses about the original data. Managing this trade-off is central to the challenge of synthetic data generation.

In the present work, we consider machine learning and probabilistic approaches to synthetic data generation and how they compare in terms of classical and original criteria reflecting the latter dimensions of interest (fidelity, utility, privacy). We focus in particular on probabilistic approaches that model the joint distribution of moderate-dimensional multivariate data via disentangling marginal and flexible dependence modeling. Our experimental results demonstrate in particular that copula-based methods achieve a sound balance across all three dimensions while remaining interpretable and providing an explicit probabilistic model. In turn, the original evaluation criteria pave the way for future research. 

One classical approach to address privacy concern explores differentially private variations of these complex methods \cite{azadmanesh2023adam, jordon2018pate}, typically by introducing noise into the learning process. This approach provides mathematical guarantees that the information gained about an individual included in the data set remains negligible \cite{dwork2006differential}. However, a lot of noise must typically be added to achieve a strong privacy guarantee, which results in insufficient synthetic data utility \cite{ganev2022robin}.  We approach this trade-off differently by using conceptually simple statistical models, which are less prone to overfitting, to generate synthetic data without having to add noise. We empirically evaluate if the generated synthetic datasets maintain sufficient privacy for our applications while guaranteeing high statistical utility.

Copulas offer a natural approach to learn joint distributions from data by decoupling the modelling of marginal distributions versus of the dependence structure  \cite{sklar1959fonctions}. This allows modelling multivariate data with various marginal distributions, as typical in medicine, and results in a flexible, interpretable method for synthetic data generation.

Previous works have utilized copulas for synthetic data generation \cite{benali2021mtcopula, houssou2022generation, li2014differentially}, but they were limited to Gaussian and Student $t$ copulas. Recently, vine copulas have also been explored for high-dimensional synthetic data generation \cite{sun2019learning}, for example in a randomized controlled trial setting \cite{petrakos2025framework}, and for data augmentation \cite{meyer2021copula}, though their interpretation can be more challenging. %Finally, another way to learn copula in a fashion similar to normalizing flows was proposed in  Kamthe \cite{kamthe2021copula}.

In this work, we specifically leverage the Gaussian Mixture Copula Model (GMCM) \cite{tewari2011parametric}, which is able to model multi-modal dependencies and which, to our knowledge, has not been explored for synthetic data generation.

\begin{figure}
    \centering
    \includegraphics[width=\columnwidth]{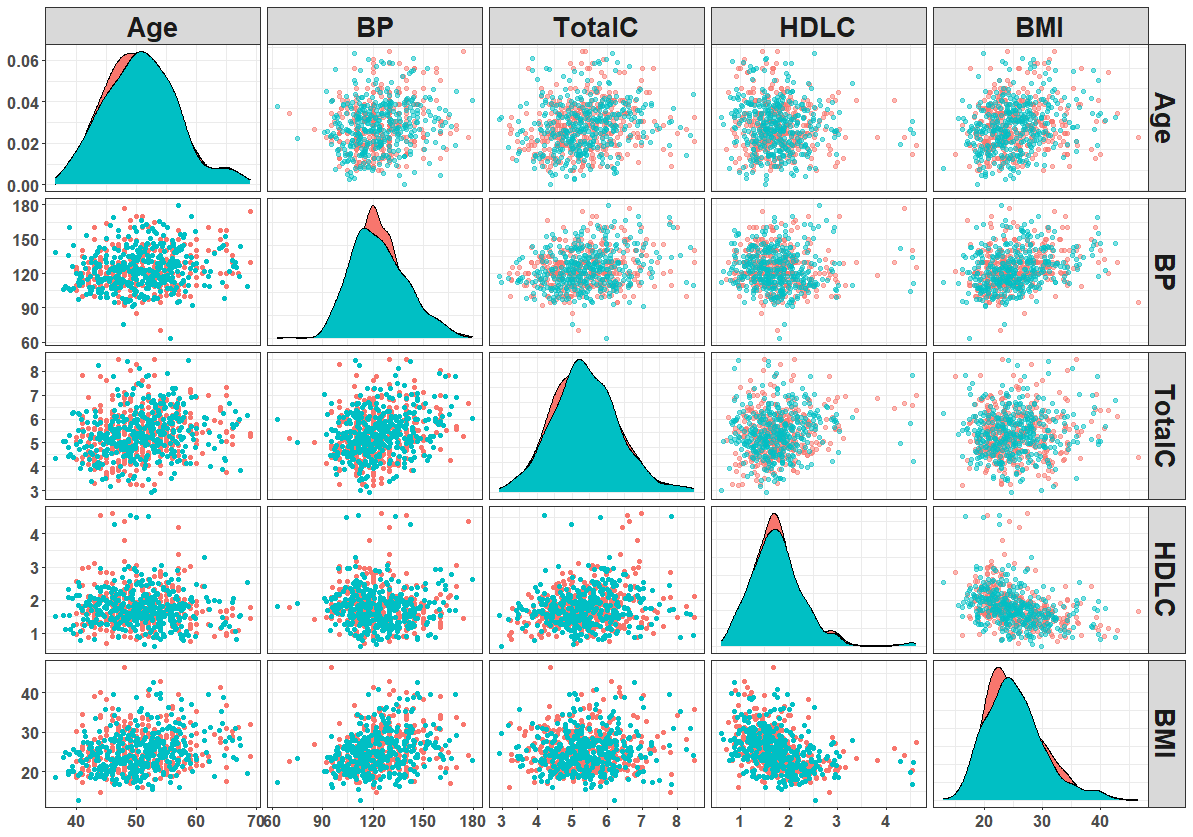}
    \caption{Synthetic data generated with the GMCM (blue) vs original data (red) from the CIMBOLIC dataset.}
    \label{fig::intro_example}
\end{figure}

We present a comprehensive framework for synthetic data generation using copulas. The approach is particularly valuable for datasets of moderate dimensions and a limited number of observations, a common scenario in medical research. While the framework is primarily aimed at continuous variables, it can also be used in the case of discrete and mixed variables. We consider the GMCM as our main proposal, alongside vine copulas and the Gaussian copula as competitors. As main example, we consider a dataset from the Department of Obstetrics and Gynecology of the University Hospital (Inselspital) in Bern, containing both personal and clinical information for 359 female patients aged 40 to 69. This dataset has been collected in the framework of the CIMBOLIC study \cite{friederichsen2020effect} and is publicly available on Zenodo \cite{faul2024data}. As shown in Figure \ref{fig::intro_example}, our method using GMCM is able to generate realistic synthetic samples which preserve marginal distributions as well as bi-variate dependence measures.

This main example focuses more specifically on an application of synthetic patient data generation to the problem of systematically missing inputs in medical risk calculators. We argue that metrics for synthetic data evaluation should be tailored to each specific situation and especially the intended applications of synthetic data generation. We recall the main evaluation objectives — \textit{fidelity}, \textit{utility}, and \textit{privacy} — and for each category we consider metrics adapted to our main example. In particular, for utility we evaluate how useful our synthetic data is for risk prediction with systematically missing inputs. For privacy, we adapt a metric simulating an attack scenario where the attacker aims to derive the cardiovascular risk of a patient, and we develop a new permutation test that compares the closest distances within real datasets to those between real and synthetic datasets — the premise being that if both datasets originate from the same distribution, these distance vectors should share similar distributions, otherwise there may be a lack of privacy protection.

Our results provide a use-inspired perspective on the fact that copula-based methods achieve an effective trade-off between fidelity, utility, and privacy by capturing the distributions of tabular datasets without overfitting. Additionally, we explore the \textit{fidelity} performance of our synthetic data generation method in other medical applications.

Furthermore, copula-based approaches for synthetic data generation offer several advantages over alternative methods. First, they yield explicit probabilistic models of the joint data distribution, enabling straightforward sampling as well as the analytical or numerical computation of quantities of interest such as marginal distributions, moments, and tail probabilities. This stands in contrast to implicit generative models such as GANs or VAEs, which provide no direct access to the underlying density. Second, the copula families investigated in this paper— including the GMCM — have closed-form conditional distributions \cite{faul2024easy}, making it possible to perform exact conditioning on observed variables without resorting to approximate inference. This can be particularly valuable in applications where one wishes to generate synthetic samples under known constraints or to impute missing values.

The key contributions of this paper are as follows:

\begin{itemize}
\item We propose a simple, interpretable framework for tabular synthetic data generation using copulas, which provides an explicit probabilistic model of the data. In particular, we introduce the Gaussian Mixture Copula Model (GMCM) for synthetic data generation for the first time, extending the existing copula-based literature beyond Gaussian, Student $t$, and vine copulas to a model capable of capturing multi-modal dependency structures.
\item Taking a use-inspired perspective, we select and adapt evaluation metrics for \textit{fidelity}, \textit{utility}, and \textit{privacy} tailored to our main application — the problem of systematically missing inputs in medical risk calculators. For \textit{privacy} in particular, we design an application-specific attack scenario and develop a permutation test based on nearest-neighbour distances.
\item On our main example and several additional medical datasets, we empirically demonstrate that copula-based methods attain a sound balance across all three evaluation dimensions of \textit{fidelity}, \textit{utility}, and \textit{privacy}, without requiring noise injection or differential privacy mechanisms, a property we attribute to their relative simplicity and natural resistance to overfitting.
\end{itemize}

The paper is organized as follows: Section \ref{sec:methodologie} introduces our general methodology for copula-based tabular synthetic data generation and present the GMCM. %Section \ref{sec::GMCM} focuses on the GMCM and discuss its properties. 
Section \ref{sec::application}  proposes an application of synthetic patient data generation to the problem of systematically missing inputs in medical risk calculators and discusses the evaluation metrics, including new proposals tailored for this task. Section \ref{sec:add_experiments} presents additional experimental results on various medical data-sets and Section \ref{sec:discussion_synthetic} discusses the advantages of our method and proposes further perspectives.

\section{From copulas to synthetic data}
\label{sec:methodologie}
Given a dataset $\mathcal{D}=(\mathbf{x}_{1},\ldots,\mathbf{x}_{n})$ assumed to be independent realizations of a random variable $\mathbf{X}$ in $\mathbb{R}^{d}$, our objective is to develop a generative model capable of sampling from this multivariate distribution.

Copula functions are cumulative distribution function (CDF) with
uniform marginals. They serve as a powerful tool for modeling the dependence structure of multivariate random variables.

Following Sklar's theorem \cite{sklar1959fonctions}, a multivariate distribution function can be modeled separately by its marginals and its copula where the copula characterizes the dependence structure between the variables.

This allows large flexibility in constructing statistical models by considering, separately, the univariate behavior of the components of a random vector and their dependence properties captured by some copulas.

Copulas are particularly useful for modeling scenarios where different types of distributions are needed for each marginal (for example multi-modal or heavy-tailed distributions) which is often the case in medical applications, providing more flexibility than classical multivariate distribution functions such as the multivariate Gaussian distributions, where every marginal distribution is required to be itself normally distributed. %For instance, in some cases, marginal distributions may exhibit different behaviors, such as varying symmetry or tail properties, which cannot be captured by a Gaussian distribution.
%Additional details about copula theory can be found in Chapter \ref{cimbolic}. 

\vspace{1em}

\textbf{Examples:}
\begin{itemize}

 \item \textbf{Gaussian Copula}: The Gaussian copula, which is the copula implicit in a multivariate Gaussian distribution, is probably the most widely used dependence structure. Its density can be expressed by:
$$c_{P}(\mathbf{u}) = \frac{\phi_{P}(\Phi^{-1}(u_{1}),\ldots,\Phi^{-1}(u_{d}))}{\prod_{i=1}^{d} \phi(\Phi^{-1}(u_{i}))} ,~ 0 \leq u_{1},\ldots,u_{d} \leq 1,$$ where $\phi$ and $\Phi$ are respectively the pdf and cdf of the univariate standard normal distribution and $\phi_{P}$ is the pdf of the multivariate $\mathcal{N}_{d}(\mathbf{0},\mathbf{P})$ distribution with $\mathbf{P}$ a correlation matrix.

\vspace{1em}

 \item \textbf{Student $t$ Copula}: The Student $t$ copula is the copula implicit in a multivariate Student's $t$-distribution. Contrary to the Gaussian copula, it allows for larger tails. Its density can be expressed by:
$$c_{P,\nu}(\mathbf{u}) = \frac{T_{\nu,P}(T_{\nu}^{-1}(u_{1}),\ldots,T_{\nu}^{-1}(u_{d}))}{\prod_{i=1}^{d} t_{\nu}(T_{\nu}^{-1}(u_{i}))},~ 0 \leq u_{1},\ldots,u_{d} \leq 1,$$ where $t_{\nu}$ and $T_{\nu}$ represent respectively the marginal cdf and pdf of a univariate Student's $t$-distribution with $\nu$ degrees of freedom and $T_{\nu,P}$ represent the joint pdf of a multivariate Student's $t$-distribution with scale matrix $P$ and $\nu$ degrees of freedom. 

\vspace{1em}

 \item \textbf{Gaussian Mixture Copula Model (GMCM)}: The Gaussian mixture copula (GMCM) \cite{tewari2011parametric} is the copula implicit in a Gaussian mixture model (GMM). The density of a GMM with $K$ components is given by:
\begin{equation*}
    \psi(x_{1},\ldots,x_{d}; \theta) = \sum_{k=1}^{K} \alpha_{k} \phi(x_{1},\ldots,x_{d};\mu_{k},\Sigma_{k}),~ \mathbf{x}= (x_{1},\ldots,x_{d}) \in \mathbb{R}^{d},
\end{equation*}
with $\alpha_{k} \geq 0,~\forall k \in \{1,\ldots,K\}$ such that $\sum_{k=1}^{K} \alpha_{k}=1$, the mixing weights and $\mu_{k},~\Sigma_{k}$ the mean and covariance matrix of the $k$-th component. $\theta = \{ \alpha_{i},\mu_{i},\Sigma_{i}\}_{i=1}^{K}$ is the set of all parameters and $\phi$ the pdf of a Gaussian distribution. The density of a GMCM with parameters $\theta$ is given by:
    \begin{equation*}
    \label{eq::density_gmcm}
        c_{gmc}(u_{1},\ldots,u_{d};\theta)=\frac{\psi(\Psi^{-1}_{1}(u_{1}),\ldots,\Psi^{-1}_{d}(u_{d}))}{\prod_{i=1}^{d} \psi_{i}(\Psi_{i}^{-1}(u_{i}))}, ~ 0 \leq u_{1},\ldots,u_{d} \leq 1,
    \end{equation*}
    where $\psi_{j}$ and $\Psi_{j}^{-1}$ denote the marginal density and inverse distribution function of the GMM along the $j$-th dimension .
The GMCM is a flexible model that captures multi-modality and asymmetry in the dependence structure, while also allowing for the capture of higher-order dependencies. It can handle both asymptotic dependent and asymptotic independent extremal dependence structures, which turns out to be an usual property for copulas in dimension larger than \(2\) \cite{andre2025gaussian}. Figure \ref{fig::different_copulas} demonstrates the flexibility of GMCM samples in capturing multimodal dependence structures, contrary Gaussian and Student $t$ copulas. %Moreover, the GMCM is also highly effective for clustering tasks. %, as detailed in Section \ref{sec:clustering} of the Appendix.

\begin{figure}
    \centering
    \includegraphics[width=0.9\columnwidth]{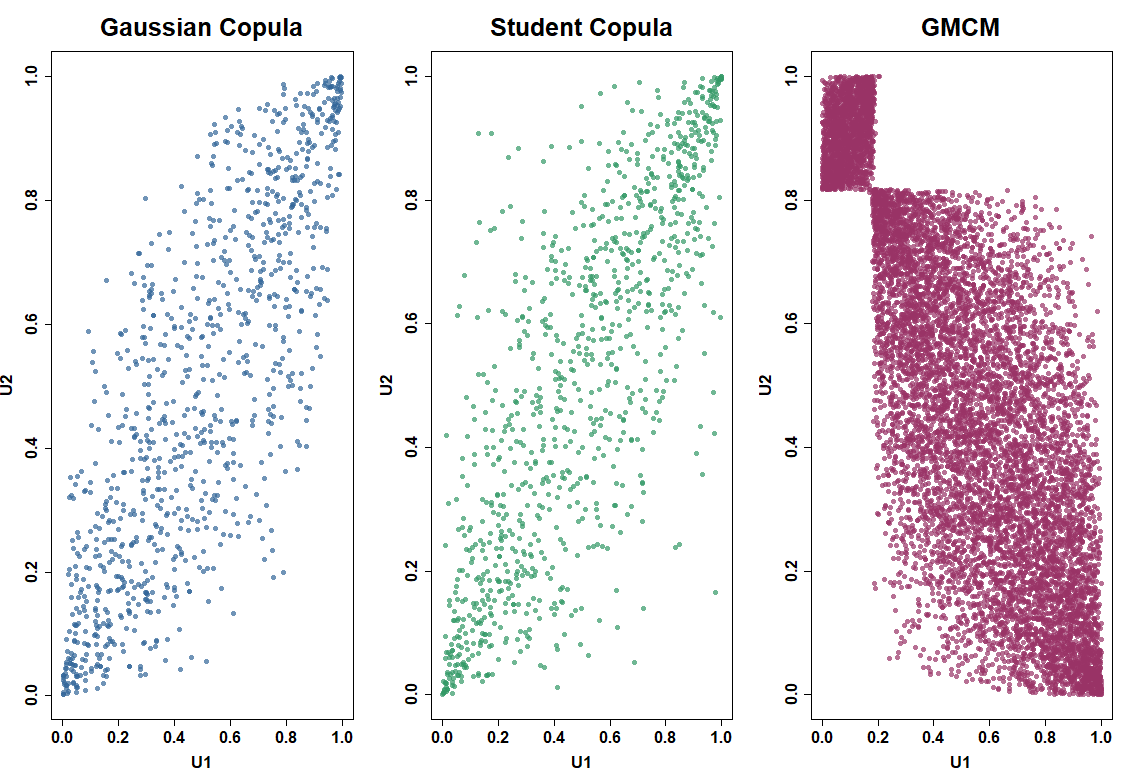}
    \caption{Samples from a Gaussian copula, a Student t copula and a Gaussian Mixture copula.}
    \label{fig::different_copulas}
\end{figure}

\vspace{1em}

 \item \textbf{Vine Copulas}: In higher dimensions, a more flexible dependence structure can be achieved through a series of simple building blocks known as pair-copulas, leading to the development of vine copulas \cite{aas2009pair}. Vine copula models construct multivariate copulas using only bivariate copulas as building blocks. The idea is to use conditioning to find a factorization of the joint density. For example, we can illustrate this for $d=3$:
$$f(x_{1},x_{2},x_{3}) = f_{1}(x_{1}) f_{2|1}(x_{2}|x_{1}) f_{3|1,2}(x_{3}|x_{1},x_{2}),$$ 
where $f_{j|G}$ is the conditional density of $X_{j}$ given variables in $G$. Using Sklar's theorem \cite{sklar1959fonctions}, we have that:
\begin{eqnarray*}
f(x_{1},x_{2},x_{3})  & = & c_{13;2}(F_{1|2}(x_{1}|x_{2}),F_{3|2}(x_{3}|x_{2});x_{2})\\ &&{}  \times c_{32}(F_{2}(x_{2}),F_{3}(x_{3}))  \\
& & {} \times c_{12}(F_{1}(x_{1}),F_{2}(x_{2})) f_{1}(x_{1}) f_{2}(x_{2}) f_{3}(x_{3}),
\end{eqnarray*}
where $c_{13;2}(\cdot,\cdot;x_{2})$ represents the copula density associated with the conditional distribution of $(X_1, X_3)$ given $X_2 = x_2$, $F_{j|G}$ is the conditional distribution function of $X_{j}$ given variables in $G$. There are various types of vine copulas, each defined by specific constraints on the factorization. Efficient fitting algorithms have been developed for these copulas, and a detailed review is available in Czado and al. \cite{czado2022vine}.
\end{itemize}

\vspace{1em}

\begin{remark}
    The Gaussian copula, Student $t$ copula and GMCM enjoys the properties of stability by marginalization and conditioning, which is practical for predictions tasks \cite{faul2024easy}.

\end{remark}

\vspace{1em}

\begin{figure*}
    \centering
    \includegraphics[width=1.05\textwidth]{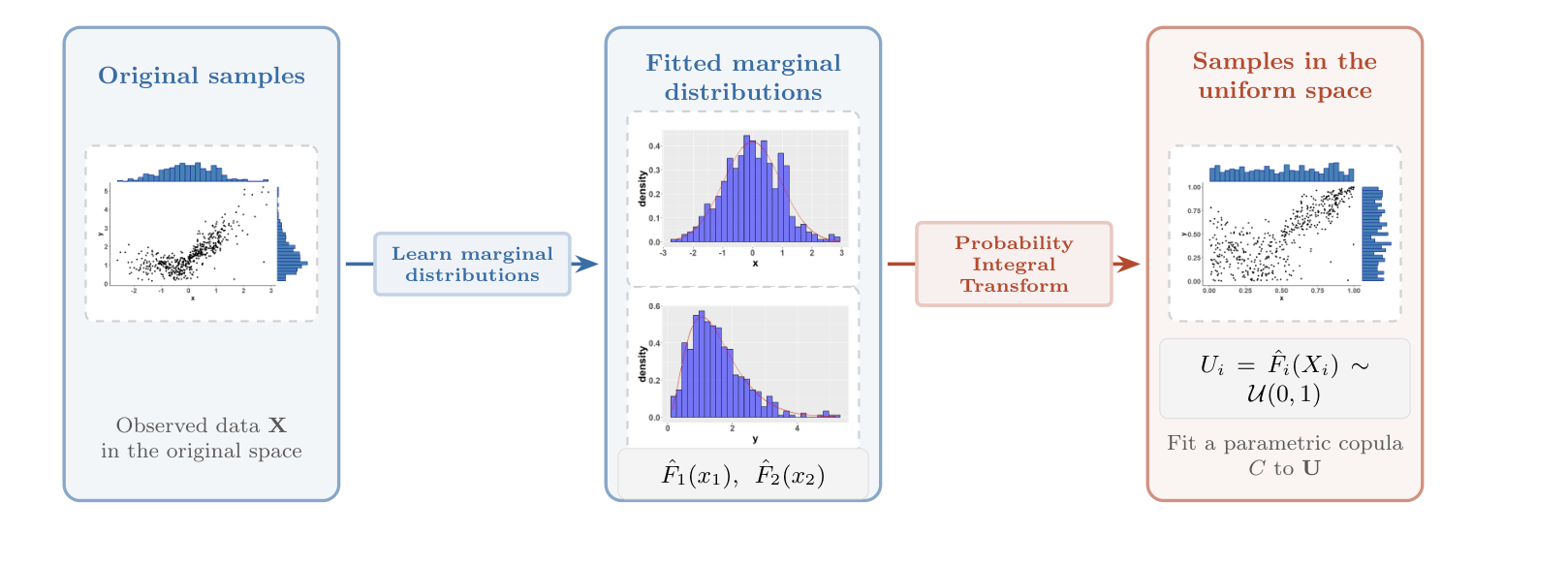}
    \caption{Illustration of the copula fitting procedure. \textbf{Step~1:} the original data $\mathbf{X}$ is observed. \textbf{Step~2:} the marginal distributions $\hat{F}_1, \hat{F}_2$ are estimated from the data. \textbf{Step~3:} the Probability Integral Transform $U_i = \hat{F}_i(X_i)$ maps approximately each variable to the uniform scale $[0,1]$; a parametric copula $C$  is then fitted to $\mathbf{U}$.}
    %\caption{Workflow for fitting a parametric copula on $2$-dimensional example.}
    \label{fig:workflow_copula}
\end{figure*}

\textbf{Copulas learning:} Fitting copulas to data is usually performed in a two-stages procedure.  First, the marginal distributions, which are typically unknown, are estimated from data. Both parametric and non-parametric methods are available for this task. Then the data can be transformed to a latent space, where the marginals are approximately uniform, by using the Probability integral transform. Utilizing Sklar's theorem \cite{sklar1959fonctions}, the copula can be inferred on this uniform space. In this work, we assume that the copula belongs to a parametric family  $\mathcal{C}= \{ c_{\theta},~\theta \in \Theta\}$,  such as the Gaussian, Student or Gaussian mixture copulas introduced earlier, and we aim to estimate the parameter $\theta$. Several methods are available for this purpose, including maximum likelihood estimation, inversion of Kendall's tau \cite{joe1996estimation}, and minimization of divergence measures such as the maximum mean discrepancy \cite{alquier2023estimation}.  Fitting of a GMCM can be challenging as detailed in the Section \ref{sec::fitting_gmcm}. The general workflow for fitting copulas to continuous data is represented on a $2$-dimensional example in Figure \ref{fig:workflow_copula}. 

\vspace{1em}

\textbf{Discrete variables:} While copulas are more straightforward to use with continuous variables, a common approach to handle discrete variables is to transform them into continuous variables, allowing for the application of copula fitting procedures typically used for continuous data. After fitting the copula model, samples can be generated and converted back to their original discrete form. Various transformations have been proposed in the literature; in this work we will use truncated normal distributions, the detailed method is described in Appendix \ref{sec::discrete}. Another approach to handling discrete variables is the extended rank likelihood method (ERL) proposed by Hoff \cite{d2007extending}. This method eliminates the need for a two-stage estimation procedure and instead uses an approximate sampling technique, such as the Metropolis-Hastings algorithm.

\vspace{1em}
%\subsection{Generating synthetic data}

\textbf{Generating synthetic data:} Once the copula and marginal distributions have been estimated, we have access to a copula generator and we can repeatedly sample a synthetic dataset of arbitrary size $m$. 
First, observations can be generated in a space with uniform marginals by sampling $m$ points from the fitted copula, $\mathbf{u}_{i} = (u^{(s)}_{i1},\ldots,u^{(s)}_{id}), ~\forall i \in  \{1,\ldots,m\}$. 
Then by applying the inverse probability integral transform to each fitted marginal, we obtain $m$ synthetic samples: $$\mathbf{x}^{(s)}_{i} = (x^{(s)}_{i1},\ldots,x^{(s)}_{id}) = \left(\hat{F}^{-1}_{1}(u^{(s)}_{i1}),\ldots, \hat{F}^{-1}_{d}(u^{(s)}_{id})\right),~\forall i \in  \{1,\ldots,m\},$$ with the desired marginal distributions.

\begin{figure*}[htbp]
    \centering
    \begin{minipage}[b]{0.48\linewidth}
        \includegraphics[width=\linewidth]{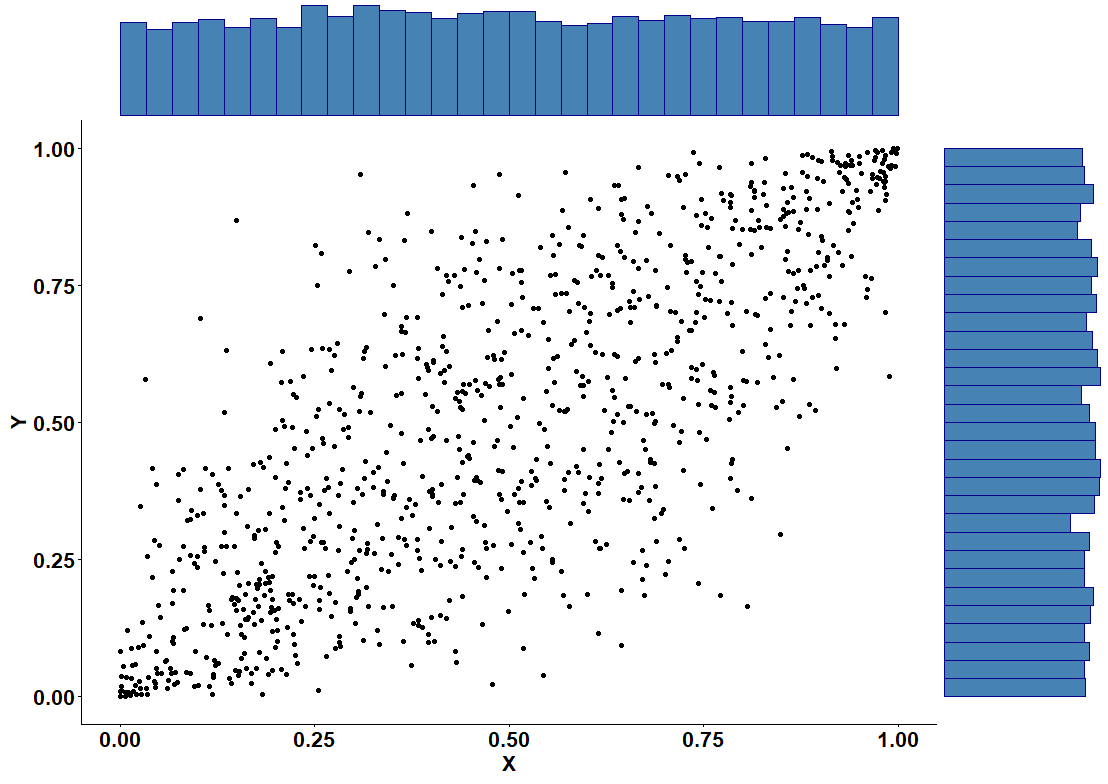}
        
        \label{fig:subfig1}
    \end{minipage}
    \hfill
    \begin{minipage}[b]{0.48\linewidth}
        \includegraphics[width=\linewidth]{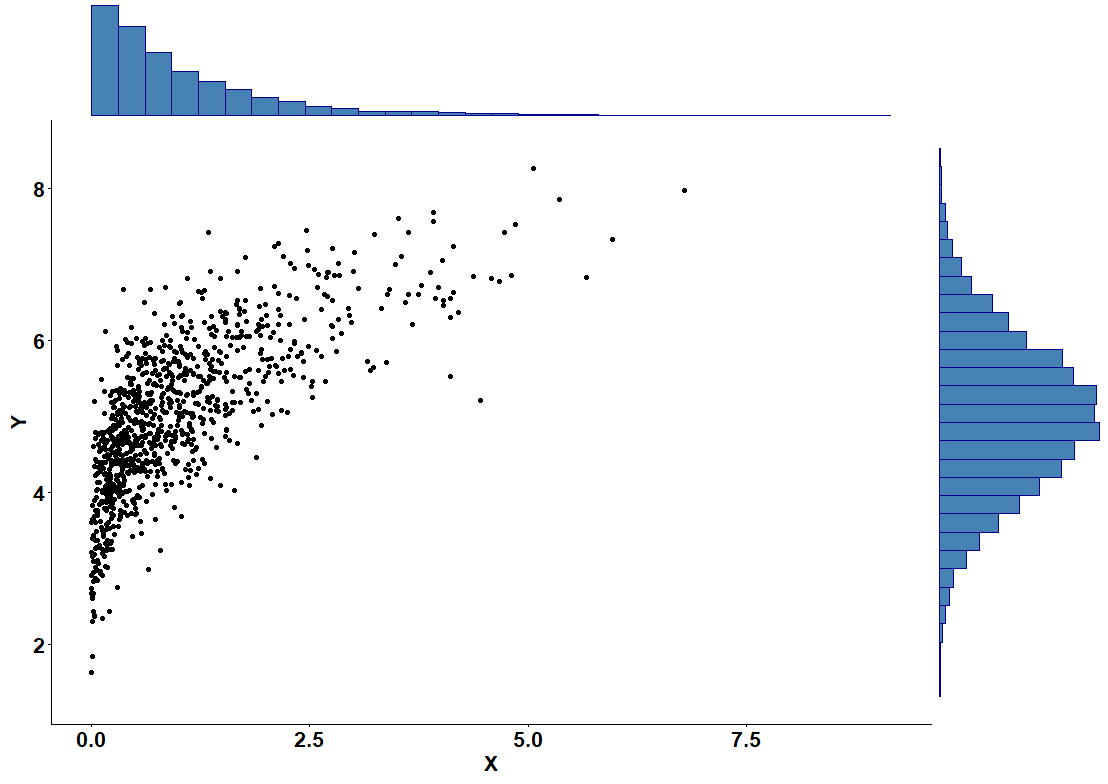}
        
        \label{fig:subfig2}
    \end{minipage}
    \caption{Samples from a Gaussian copula with $\rho=0.8$ in the uniform space (left) and after transforming the marginals (right).}
    \label{fig:copula_generation}
\end{figure*}

Figure \ref{fig:copula_generation} illustrates the process of generating points using a bivariate Gaussian copula with a correlation parameter ($\rho = 0.8$). Initially, points are drawn from the copula in the uniform, and subsequently, the marginal distributions are transformed into Gaussian and Gamma distributions, respectively.

\vspace{1em}

\section{Generating and evaluating synthetic datasets: a case study}
\label{sec::application}

\subsection{General principles for empirically evaluating synthetic datasets}
\label{sec::evaluation}

Evaluating the quality of synthetic datasets is a challenging task due to the the inherent trade-off between statistical similarity and privacy. Additionally, the extensive range of evaluation metrics available can make it challenging to select the most appropriate one to use. 

In this work, we focus on the empirical evaluation of synthetic datasets and we claim that the evaluation should align with the specific purpose for which the synthetic data is generated, necessitating the metrics to be adapted for each case.

We highlight the three key attributes of high-quality synthetic data: \textit{fidelity}, \textit{utility}, and \textit{privacy}.

\begin{itemize}

\item \textbf{Fidelity}: From a high-level standpoint, \textit{fidelity} refers to how closely the synthetic data statistically resembles the real data. It is crucial to define which aspects of the original data should be preserved.

\item \textbf{Utility}: The \textit{utility} of synthetic data is evaluated based on its performance in practical applications, such as downstream regression or classification tasks. We employ the Train on Synthetic Test on Real (TSTR) paradigm, described later in this section, to assess the \textit{utility} of synthetic datasets.

\item \textbf{Privacy}: \textit{Privacy} assesses how much information the synthetic data discloses about the original data. The  goal is to prevent replication of individuals from the real data and protect sensitive attributes from potential attacks.

\end{itemize}
With the general measures for evaluating synthetic datasets established, and in line with our view that evaluation should be tailored to the specific application, we will later apply these principles to the specific test case presented next.

\subsection{Description of the case study}
\label{sec:case_study}

We address the challenge of systematic missing values in medical risk calculators, focusing on the European Systematic Coronary Risk Evaluation 2 (SCORE2) \cite{esc2021score2}. SCORE2 estimates the 10-year risk of fatal and non-fatal cardiovascular events across Europe and has been calibrated for Switzerland, a low-risk region. The calculator requires age, blood lipid levels (HDL and total cholesterol), systolic blood pressure, smoking status, and personal history of diabetes mellitus. A separate calculator, SCORE2-Diabetes, exists for patients with type 2 diabetes but is not considered in this work.

Women in midlife represent a particularly relevant population for cardiovascular risk prediction. The menopausal transition is associated with adverse changes in body composition, lipid profile, blood pressure, and glucose metabolism, collectively contributing to a substantial increase in cardiovascular risk. Consequently, cardiovascular risk assessment is increasingly recommended during and after the menopausal transition, as cardiovascular disease remains the leading cause of mortality among women. However, complete laboratory measurements are not always available in first consultations, workplace screening programs, or digital health interventions—settings in which risk calculators are most likely to be deployed.

Following the setting described in M\"{u}hlemann et al.\ \cite{muhlemann2025comparing}, we assume that three clinical inputs required by SCORE2---systolic blood pressure, total cholesterol, and HDL cholesterol---are systematically missing, as they require clinical or laboratory measurements that can be costly and time-consuming to obtain. We acknowledge that the systematic missingness assumption is more realistic for the lipid values than for systolic blood pressure, which can be measured cheaply at home or in a pharmacy. 

Our objective is to obtain accurate probabilistic predictions of the CVD risk when these clinical variables are missing by using statistical methods to substitute them with self-acquired information such as age and body mass index (BMI). Our approach involves generating probabilistic predictions for the missing inputs in the risk calculator and propagating this uncertainty to provide a probabilistic risk prediction.

We use data collected at the Department of Obstetrics and Gynecology of the University Hospital (Inselspital) in Bern, Switzerland, comprising 359 women aged 40 to 69. The dataset originates from the CIMBOLIC study \cite{friederichsen2020effect}, a retrospective cohort study of women with menopausal symptoms that examined the influence of \textit{Cimicifuga racemosa} extract (CR, Ze~450) and menopausal hormone therapy (MHT) on metabolic parameters and body weight. The study was approved by the Bernese Cantonal Ethics Committee (Kantonale Ethikkommission Bern, KEK Switzerland, No.\ 2016-01179) and the dataset is publicly available on Zenodo \cite{faul2024data}.

\subsection{Synthetic data generators}

We implemented the Gaussian Mixture Copula model (GMCM) with non-parametric fitting of the marginals using kernel density estimation (KDE) and set the number of clusters to three. We compared it with a Gaussian copula synthesizer (GC), a conditional tabular generative adversarial network synthesizer (CTGAN) \cite{xu2019modeling}—both available in the \textit{Python} package SDV \cite{patki2016synthetic} —and an adversarial random forests-based generator (ARF) \cite{watson2023adversarial}. Details about ARF and CTGAN methods can be found in Appendix \ref{sec::other_generators}. Additionally, we included a simple benchmark modeling only the marginal distributions with kernel density estimation (MARG). Finally, we implemented a vine copula model (Vines) \cite{aas2009pair}, with marginals fitted by kernel density estimation.

We generated \(100\) datasets, each containing \(1,000\) elements, for every method. All results are reported as the mean \(\pm\) standard deviation, averaged over these \(100\) synthetic datasets. More details about the experiments can be found in Appendix \ref{sec:experimental_details_synthetic}.

\subsection{Empirical evaluation of synthetic datasets for our case study}
\label{sec:results}

Having defined the general principles for evaluating synthetic datasets in Section~\ref{sec::evaluation}, we now specify the desiderata for the synthetic data used in our case study. With respect to \textit{fidelity}, we aim to preserve the marginal distributions of each input of the risk calculator, as well as the marginal distribution of the SCORE2 risk. We also seek to preserve the dependence structure among variables—that is, the joint distribution—as closely as possible. With respect to \textit{utility}, we require the synthetic datasets to replace the real data for a probabilistic prediction task of the SCORE2 risk when clinical measurements are \textit{systematically missing}. Finally, with respect to \textit{privacy}, we treat the SCORE2 risk as a sensitive attribute that should not be inferable, and we aim to prevent approximate duplication of real patients in the synthetic dataset. For each of these desiderata, we introduce specific metrics and present empirical results.

\vspace{1em}

1)  \textbf{Marginal distributions: }

\vspace{1em}

To show that the marginal distributions are preserved, a first idea is to  visualize the empirical distributions of the real versus a synthetic dataset. In Figure \ref{fig::intro_example}, we see that the resulting synthetic data closely resembles the original data. For illustrative purposes, we have included only the continuous variables in this representation. 

Additionally, we compare the marginal distributions of the original dataset and the generated synthetic datasets using basic statistics such as mean and standard deviation. %To evaluate whether the marginal distributions differ significantly between real and synthetic datasets, we employ statistical tests: the Student's t-test for continuous variables and the Pearson Chi-square test for binary variables. 
The results are presented in Table \ref{tab:marginals} of Appendix \ref{sec::marginal_eval}.

Given that the copula-based approaches GMCM and Vines, as well as MARG, employ non-parametric estimation of the marginals using KDE, they are anticipated to effectively capture the marginal distributions, a performance confirmed by Table \ref{tab:marginals}, with ARF also showing strong results. However, the findings indicate that CTGAN produces synthetic samples that significantly differ from the real samples across all marginal variables.

\vspace{1em}

2)  \textbf{Joint distribution:}

\vspace{1em}

Above, we only evaluate the marginal distribution mismatch for each feature. In this part, we
show how well the generated synthetic data preserves the
“joint” distribution of the original data. To measure the distance between joint distributions of the original dataset and
the synthetic datasets, we consider general statistical metrics that are often used in the synthetic data generation literature.

The similarity between generated synthetic data $\mathbf{X}^{(s)} =  ( { \mathbf{x}_1^{(s)}, \mathbf{x}_2^{(s)}, \ldots, \mathbf{x}_m^{(s)} } ) $ and the empirical training data $\mathbf{X}^{(r)} = ( { \mathbf{x}_1^{(r)}, \mathbf{x}_2^{(r)}, \ldots, \mathbf{x}_n^{(r)} } )$ can be assessed by the Maximum Mean Discrepancy (MMD) \cite{gretton2012kernel} which can be estimated by:
\begin{equation*}
    \begin{split}
        \text{MMD}^2(\mathbf{X}^{(s)}, \mathbf{X}^{(r)}) = \frac{1}{m^2} \sum_{i=1}^{m} \sum_{j=1}^{m} k(\mathbf{x}_i^{(s)}, \mathbf{x}_j^{(s)}) \\ + \frac{1}{n^2} \sum_{i=1}^{n} \sum_{j=1}^{n} k(\mathbf{x}_i^{(r)}, \mathbf{x}_j^{(r)}) \\ - \frac{2}{nm} \sum_{i=1}^{n} \sum_{j=1}^{m} k(\mathbf{x}_i^{(s)}, \mathbf{x}_j^{(r)}),
    \end{split}
\end{equation*}

where $k(\cdot, \cdot)$ is a positive definite kernel. We consider the classical isotropic Gaussian kernel between two points $\mathbf{x},\mathbf{x'} \in \mathbb{R}^d$ given by:
\begin{equation*}
k_{\text{iso}}(\mathbf{x},\mathbf{x'}) = \sigma^2 \exp \left( -\frac{1}{2\ell^2} \| \mathbf{x}-\mathbf{x'} \|^2 \right),
\end{equation*}
where $\ell > 0$ is the shared length scale chosen using the median heuristic \cite{gretton2012kernel} , and $\sigma^2$ controls the amplitude.  We will denote the resulting isotropic MMD as MMDi in the following.

We also consider an anisotropic version which replaces the scalar length scale by a direction-dependent metric. In its most general form:
\begin{equation*}
k_{\text{aniso}}(\mathbf{x},\mathbf{x'}) = \sigma^2 \exp \left( -\frac{1}{2} (\mathbf{x}-\mathbf{x'})^{\top} \Sigma^{-1} (\mathbf{x}-\mathbf{x'}) \right),
\end{equation*}

where $\Sigma$ is a $d \times d$ symmetric, positive-definite covariance matrix encoding the length-scales across all direction, estimated using the median heuristic in each dimension. We will denote the resulting anisotropic MMD as MMDa in the following.

Additionally, we consider the Inverse Kullback-Leibler divergence (IKL) and the Jensen-Shannon Divergence (JSD) between the real dataset and the synthetic dataset as other measures of performances. 

Finally, another way to evaluate synthetic data is to use Precision and Recall metrics \cite{kynkaanniemi2019improved}. Precision measures fidelity as the expected likelihood that a synthetic sample lies within the support of the real distribution, where the support is the set of possible real samples with non-zero probability density. Recall, in contrast, measures diversity as the expected likelihood that a real sample lies within the support of the synthetic distribution.

Formally, if we denote $\mathbf{X}_{r} \sim P_{r}$ and $\mathbf{X}_{s} \sim P_{s}$, random variables following respectively the distribution of the real and synthetic data distribution.

The precision and recall are defined by:
$$\text{Precision} := P = \mathbb{E}_{\mathbf{X}_{s} \sim P_{s}} (\mathbf{X}_{s} \in \text{Supp}(P_{r}))$$ and $$\text{Recall} := R =  \mathbb{E}_{\mathbf{X}_{r} \sim P_{r}} (\mathbf{X}_{r} \in \text{Supp}(P_{s})).$$
To estimate these metrics from samples of $P_{s}$ noted $(\mathbf{x}_{1}^{(s)},\ldots,\mathbf{x}_{m}^{(s)})$ and from $P_{r}$ noted $(\mathbf{x}_{1}^{(r)},\ldots,\mathbf{x}_{n}^{(r)})$, we define:
$$\hat{P} := \frac{1}{m} \sum_{j=1}^{m} BSR_{P_{r}}(\mathbf{x}_{j}^{(s)})$$
$$\hat{R} := \frac{1}{n} \sum_{i=1}^{n} BSR_{P_{s}}(\mathbf{x}_{i}^{(r)}),$$
where $$BSR_{P_{r}}(\mathbf{x}_{j}^{(s)}) =  \left\{
    \begin{array}{ll}
        1& \mbox{if } \mathbf{x}_{j}^{(s)} \in \bigcup_{i=1}^{n} \mathcal{B}(\mathbf{x}_{i}^{(r)},NND_{K}(\mathbf{x}_{i}^{(r)})) \\
        0 & \mbox{otherwise}
    \end{array}
\right.,
$$
with $\mathcal{B}(\mathbf{x},r)$ hypersphere in $\mathbb{R}^{d}$ around $\mathbf{x}$ with radius $r$ and $NND_{K}(\mathbf{x})$ the euclidean distance from $\mathbf{x}$ to the $K$-th nearest neighbors among $\{\mathbf{x}_{i}^{(r)} \}_{i=1}^{n}$ excluding itself and $BSR_{P_{s}}(\mathbf{x}_{j}^{(r)})$ is defined similarly.

However, recent work has highlighted the limitations of using Precision and Recall for the empirical evaluation of synthetic datasets and has proposed probabilistic alternatives, namely Probabilistic Precision (P-PR) and Probabilistic Recall (P-RE)~\cite{park2023probabilistic}. For these metrics, we assume that the supports of $P_{r}$ and $P_{s}$ are random variables. Given the observations $\{\mathbf{x}_{i}^{(s)}\}_{i=1}^{m}$ (resp. $\{\mathbf{x}_{i}^{(r)}\}_{i=1}^{n}$), we assume that there are
myriad of possible supports of $P_{s}$ (resp. $P_{r}$), denoted as $S_{P_{s}}$ (resp. $S_{P_{r}}$). The collection of these supports is denoted by $\mathbb{S}_{P_{s}}$ (resp. $\mathbb{S}_{P_{r}}$). The Probabilistic Precision is then defined as $$\text{P-Precision}= \mathbb{E}_{\mathbf{X}_{s} \sim P_{s}, S_{P_{r}} \in \mathbb{S}_{P_{r}}} ( \mathbf{X}_{s} \in \mathbb{S}_{P_{r}}) = \mathbb{E}_{\mathbf{X}_{s} \sim P_{s} }(\mathbb{P}(\mathbf{X}_{s} \in S_{P_{r}}))$$ and can be estimated by:
$$\text{P-Precision} \approx \frac{1}{m} \sum_{j=1}^{m} \mathbb{P}(\mathbf{x}_{j}^{(s)} \in S_{P_{r}}).$$ In a similar manner, the Probabilistic Recall is defined by:
 $$\text{P-Recall}= \mathbb{E}_{\mathbf{X}_{r}\sim P_{r}, S_{P_{s}} \in \mathbb{S}_{P_{s}}} ( \mathbf{X}_{r} \in \mathbb{S}_{P_{s}}) = \mathbb{E}_{\mathbf{X}_{r} \sim P_{r}}(\mathbb{P}(\mathbf{X}_{r}\in S_{P_{s}}))$$
and can be estimated by:
$$\text{P-Recall} \approx \frac{1}{n} \sum_{i=1}^{n} \mathbb{P}(\mathbf{x}_{i}^{(r)} \in S_{P_{s}}).$$
Additional details about the practical computations of these quantities can be found in Park and al. \cite{park2023probabilistic}.

All these metrics are implemented in the \textit{Python} package Synthcity \cite{qian2023synthcity}. The results in Table \ref{tab:cimbolic_comparison} indicate that for most metrics, GMCM, ARF, and Vines are the top-performing methods (highlighted in green). Overall, copula-based methods demonstrate performance comparable to ARF, which is renowned for its effectiveness across various applications. Note that the DCR and SensAttack metrics are introduced later in this section.
\begin{table*}
    \renewcommand{\arraystretch}{2.5}
    \resizebox{\textwidth}{!}{
        \Large
        %\scalebox{1.8}
        {
        \begin{tabular}{p{5cm}cccccccc}
            \toprule
            Generator Type & MMD (e-3) $\downarrow$ & MMDa (e-3) $\downarrow$ & JSD (e-3) $\downarrow$ & IKL (e-1) $\uparrow$ & P-PR $\uparrow$ & P-RE $\uparrow$ & DCR $\uparrow$ & SensAttack $\uparrow$\\
            \midrule
        GC & \cellcolor{gray!30}4.08 $\pm$ 0.02 & \cellcolor{red!30}31.74 $\pm$ 0.62 & \cellcolor{red!30}13.94$\pm$ 0.58 & \cellcolor{red!30}7.71 $\pm$ 0.14 & \cellcolor{red!30}0.99 $\pm$ 0.003 & \cellcolor{red!30}0.88  $\pm$ 0.02 & 0.28$\pm$ 0.006 & 0.11 $\pm$ 0.01\\
CTGAN & \cellcolor{gray!30}4.0 $\pm$ 0.007 & \cellcolor{red!30}44.59 $\pm$ 1.37 & \cellcolor{red!30}31.60 $\pm$ 0.58 & \cellcolor{red!30}7.97 $\pm$ 0.04 & \cellcolor{red!30}0.90  $\pm$ 0.009 & \cellcolor{red!30}0.98  $\pm$ 0.006 & 0.59 $\pm$ 0.01 & 0.48 $ \pm$ 0.01\\
GMCM & \cellcolor{green!30}3.96 $\pm$ 0.01 & \cellcolor{green!30}3.86 $\pm$ 0.41 & \cellcolor{green!30}3.38$\pm$ 0.39 & \cellcolor{green!30}9.90$\pm$ 0.03 & \cellcolor{green!30}0.97  $\pm$ 0.004 & \cellcolor{green!30}0.98  $\pm$ 0.005 & 0.31 $\pm$ 0.007 &  0.14 $\pm$ 0.004\\
MARG & \cellcolor{gray!30}3.97 $\pm$ 0.01 & \cellcolor{red!30}7.34 $\pm$ 0.54 & \cellcolor{green!30}3.35 $\pm$ 0.41 & \cellcolor{green!30}9.90 $\pm$ 0.037 & \cellcolor{green!30}0.97  $\pm$ 0.005 & \cellcolor{green!30}0.98  $\pm$ 0.005 & 0.34 $\pm$ 0.009 & 0.32 $\pm$ 0.02\\
ARF & \cellcolor{green!30}3.93 $\pm$ 0.01 & \cellcolor{green!30}3.10 $\pm$ 0.39& \cellcolor{gray!30}3.93 $\pm$ 0.44 & \cellcolor{green!30}9.84 $\pm$ 0.07 & \cellcolor{green!30}0.97 $\pm$ 0.004 & \cellcolor{green!30}0.99 $\pm$ 0.004 & 0.30 $\pm$ 0.008 & 0.11 $\pm$ 0.004\\
Vines & \cellcolor{green!30}3.96 $\pm$ 0.01 & \cellcolor{green!30}3.60 $\pm$ 0.33 & \cellcolor{green!30}3.34 $\pm$ 0.44 & \cellcolor{green!30}9.91 $\pm$ 0.03& \cellcolor{green!30}0.97 $\pm$ 0.004 & \cellcolor{green!30}0.98  $\pm$ 0.005 & 0.31 $\pm$ 0.008 & 0.13 $\pm$0.01\\
            \bottomrule
        \end{tabular}
        }
    }
    \vspace{1em}
    \caption{Experiment Results Over the Cimbolic dataset. For each evaluation metric, the down (up) arrow means the lower (higher) the better respectively. The unit of value for each column is stated in the bracket right after the corresponding variable name in the table. Bold indicates the best performance. All results are reported as the mean \(\pm\) std, averaged over 100 synthetic dataset.}
    \label{tab:cimbolic_comparison}
\end{table*}

\newpage

3) \textbf{Probabilistic prediction:}

\begin{figure}[h]
    \centering
    \includegraphics[width= 0.7\columnwidth]{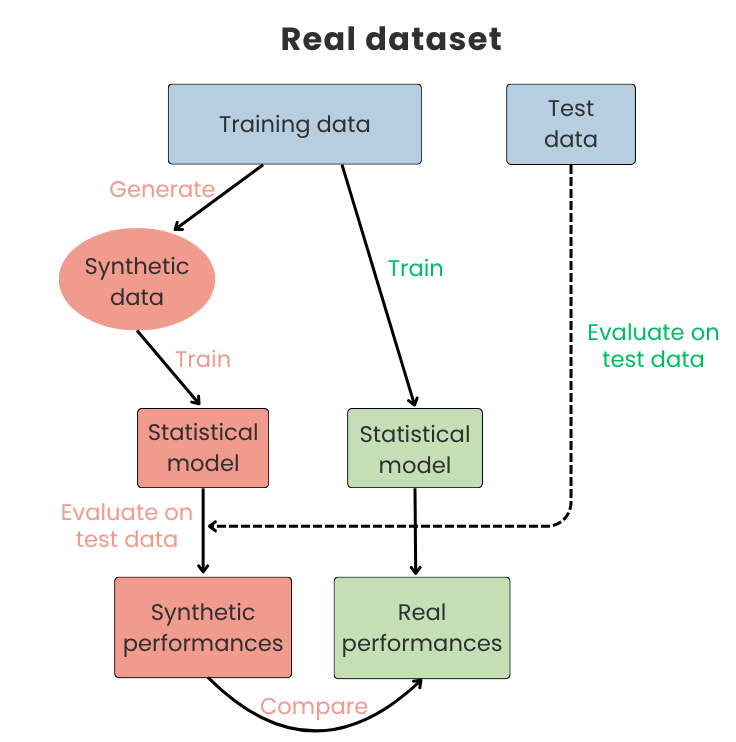}
    \caption{Train on Synthetic Test on Real evaluation framework}% \cite{WinNT}}
    \label{fig::synthetic_cymbolic}
\end{figure}

We measure the utility of a synthetic data set by assessing the similarity of results for specific analyses using both the original and the synthetic data. For high utility, we expect close similarity between the results for the same analysis
calculated from the two different sources of data. A way to measure this is commonly referred to as the Train on Synthetic Test on Real (TSTR) evaluation \cite{esteban2017real}, it is illustrated on Figure \ref{fig::synthetic_cymbolic}.

In our study, we focus on probabilistic predictions of the SCORE2 risk in scenarios where specific clinical measurements are \textit{systematically missing}. Specifically, we consider the case where Systolic Blood Pressure, Total Cholesterol, and HDL Cholesterol are \textit{systematically missing}. We employ $10$-fold cross-validation, ensuring that for each left-out fold, all clinical data is available in the training set, while these specified measurements are absent in the test set.

For this task, we consider the method introduced in M\"uhlemann and al. \cite{muhlemann2025comparing}, which consist in probabilistically imputing the missing variables and then propagating the uncertainty through the risk calculator. The model used for probabilistic imputation is a Gaussian copula model. To evaluate the accuracy of the probabilistic predictions of CVD risk, we use proper scoring rules \cite{gneiting2007strictly} such as the Continuous Ranked Probability Score (CRPS), the weighted CRPS (wCRPS) and the log-score (LogS).

In Table \ref{tab:TSTR}, we present the predictive performance of various synthetic data generators, averaged over multiple train-test splits and synthetic datasets. The “real data” column represents the performance of our model trained and evaluated with the real data. Our goal is for the performance of models trained on generated synthetic datasets to closely match that of the real data. The results show that the copula-based method, GMCM, Vines and ARF, aligns most closely with the performance of real data, demonstrating the effectiveness of copula-based synthetic data generation approach in predicting the SCORE2 risk, even when confronted with systematically missing inputs.

\begin{table}
\centering
\renewcommand{\arraystretch}{0.8}

{
{%
\begin{tabular}{l|c|c|c}
\hline
Generator Type & CRPS \(\downarrow\) & wCRPS \((1e{-}2)\) \(\downarrow\) & Log-Score \(\downarrow\) \\
\hline
Real  & 0.340 $\pm$ 0.031 & 5.03 $\pm$ 2.16 & 0.781 $\pm$ 0.074 \\
\hline
GC    & 0.358 $\pm$ 0.040 & 5.23 $\pm$ 2.71 & 0.926 $\pm$ 0.169 \\
CTGAN & 0.550 $\pm$ 0.042 & 8.17 $\pm$ 2.60 & 1.26 $\pm$ 0.057 \\
GMCM  & 0.346 $\pm$ 0.047 & 5.16 $\pm$ 3.23 & 0.808 $\pm$ 0.0101 \\
MARG  & 0.395 $\pm$ 0.044 & 5.77 $\pm$ 3.13 & 0.939 $\pm$ 0.098 \\
ARF   & 0.348 $\pm$ 0.036 & 5.14 $\pm$ 2.59 & 0.822 $\pm$ 0.083 \\
Vines & 0.343 $\pm$ 0.039 & 5.05 $\pm$ 2.36 & 0.797 $\pm$ 0.090 \\
\hline
\end{tabular}%
}
}

\vspace{1em}
\caption{Performances of SCORE-2 risk prediction with systematic missing values by using different synthetic dataset in the TSTR setting. Real data indicates the performances when training and testing on the real data. All results are reported as the mean \(\pm\) std, averaged over \(100\) synthetic dataset and \(20\) train-test splits.}
%\vspace{1em}
\label{tab:TSTR}
\end{table}

\newpage

4) \textbf{Sensitive attribute:}

\vspace{1em}

To evaluate the vulnerability of sensitive attributes,
we introduce a “SensAttack” metric. % to evaluate how well sensitive attributes are protected in real datasets.
 This metric simulates an attack scenario where the attacker has access to both the synthetic data and the non-sensitive attributes of real observations, with the goal of inferring the sensitive attributes of these real observations.

The metric is constructed as follows: for each record in the real dataset, we identify the $K$ nearest neighbors in the synthetic data based on non-sensitive attributes. We predict the sensitive attributes of the real record by averaging the corresponding values from these $K$ nearest synthetic neighbors. This prediction is then compared to the prediction made using the $K$ nearest neighbors from the real dataset. We use the difference in mean squared error (MSE) between predictions based on synthetic data and those based on real data as our evaluation metric. A value smaller than zero indicates that having access to the synthetic data improves the prediction of sensitive attributes compared to using real data alone, suggesting potential privacy breaches. Conversely, a positive value reflects better privacy protection of the sensitive attribute.

The hyperparameter $K$ should be set to a low value, such as $1$ or $2$, to emphasize our intention to avoid duplications in the synthetic dataset. We employ the Gower distance \cite{gower1971general} to determine the nearest neighbors, given its capability to effectively handle both continuous and categorical features.

This metric extends the \textit{Disclosure Protection} metric from the SDV package \cite{patki2016synthetic} by evaluating the risk of disclosing a sensitive attribute during a privacy attack. Unlike \textit{Disclosure Protection}, our metric accommodates numerical sensitive attributes and utilizes nearest neighbors, offering a more generalized approach.

In this application, we consider the SCORE2 risk to be the sensitive attribute even if it isn't directly observed in the data. The results are presented in the last column of Table \ref{tab:cimbolic_comparison}. Every synthetic data generation methods has a “SensAttack” score larger than $0$ which indicates a good protection of the SCORE2 risk.

\vspace{1em}

5) \textbf{No Patient Duplication:}

\vspace{1em}

Finally to assess that there is no real patient which has been (approximately) duplicated in the synthetic data, we considered the Distance to Closest Record (DCR) metric, implemented in the \textit{Python} package SDV \cite{patki2016synthetic}. The DCR metric ranges from $0$ to $1$, reflecting how closely synthetic data resembles real data compared to random data. It achieves this by independently sampling each marginal from a uniform distribution in their range, which provides optimal privacy protection. A DCR score of $1$ suggests that synthetic data maintains maximum privacy, while a score of $0$ indicates minimal privacy. For instance, a score of $0.3$ reflects that the synthetic data is $70\%$ closer to the real data than random data.

In our experiments, we set a DCR threshold of $0.1$, deeming a synthetic data generation method as private only if the synthetic data is less than $90\%$ closer to real data than randomly generated data would be. We acknowledge that this threshold may be somewhat arbitrary. In Section \ref{sec:discussion_synthetic}, we will explore other privacy assessment metrics based on permutation tests.

The results are summarized in the second to last column of Table \ref{tab:cimbolic_comparison}. All generators have a DCR larger than $10 \%$ so we will consider all of them as sufficiently private. We acknowledge that this $10\%$ threshold may seem arbitrary; we therefore propose an alternative measure based on permutation tests in Section~\ref{sec:discussion_synthetic}.

\section{Additional experiments}
\label{sec:add_experiments}
In this section, we apply our methodology for synthetic data generation on two real world medical datasets. Additional experiments on simulated data can be found in Appendix \ref{sec:synthetic_experiments}.

\subsection{Wisconsin Breast Cancer Data}
\label{sec:wisconsin}
The Breast Cancer Wisconsin dataset, sourced from the UCI Machine Learning Repository, utilizes digitized images from fine needle aspirates (FNA) of breast masses. It comprises ten features derived from \(569\) patients. For each feature, the dataset provides the mean value, extreme value (average of the three largest values), and standard error, resulting in \(30\) continuous variables. The dataset includes two types of diagnoses: benign (\(352\) patients) and malignant (\(212\) patients), suggesting the presence of two distinct clusters. For this illustration, we focus on a subset of four variables: perimeter standard error (PSE), extreme smoothness (ES), extreme concavity (EC), and extreme concave points (ECP), similarly to previous works \cite{sahin2022vine}. %This dataset has been widely utilized for clustering tasks. Given its multi-modal structure, we expect the GMCM to perform particularly well.

We synthesized $100$ datasets for each method and evaluated the generated synthetic data using statistical utility and privacy metrics, similarly to the approach in Section \ref{sec::application}.

\begin{table*}[h]
    \renewcommand{\arraystretch}{2.5}
    \resizebox{1.0\textwidth}{!}{
        %\Large
        %\scalebox{1.8}{
        \begin{tabular}{lccccccc}
            \toprule
            Generator Type & MMD (e-3) $\downarrow$ & MMDa (e-3) $\downarrow$ & JSD (e-3) $\downarrow$ & IKL $\uparrow$ & P-PR $\uparrow$ & P-RE $\uparrow$ & DCR (e-2) $\uparrow$\\
            \midrule
            GC & \cellcolor{red!30}7.01 $\pm$ 2.37 & \cellcolor{red!30}26.63 $\pm$ 1.54 & \cellcolor{red!30}12.75 $\pm$ 0.59 & \cellcolor{red!30}0.77 $\pm$ 0.01 & \cellcolor{green!30}0.98 $\pm$ 0.003 & \cellcolor{green!30}0.97 $\pm$ 0.005 & 0.13$\pm$ 0.003 \\
            CTGAN & \cellcolor{red!30}14.69 $\pm$ 1.59 & \cellcolor{red!30}79.37 $\pm$ 2.87 & \cellcolor{red!30}24.81 $\pm$ 0.53 & \cellcolor{red!30}80.22 $\pm$ 0.08 & \cellcolor{red!30}0.80 $\pm$ 0.01 & \cellcolor{red!30}0.98 $\pm$ 0.006 & 0.27 $\pm$ 0.01 \\
            GMCM & \cellcolor{green!30}1.86 $\pm$ 0.89 & \cellcolor{green!30}2.63 $\pm$ 0.69 & \cellcolor{gray!30}4.64$\pm$ 0.51 & \cellcolor{green!30}0.98 $\pm$ 0.006 & \cellcolor{green!30}0.97 $\pm$ 0.005 & \cellcolor{green!30}0.99 $\pm$ 0.004 & 0.13 $\pm$ 0.004 \\
            MARG & \cellcolor{red!30}5.73 $\pm$ 1.27 & \cellcolor{red!30}42.76 $\pm$ 2.48 & \cellcolor{gray!30}4.71 $\pm$ 0.50 & \cellcolor{green!30}0.98 $\pm$ 0.007 & \cellcolor{red!30}0.82 $\pm$ 0.01 & \cellcolor{red!30}0.99 $\pm$ 0.005 & 0.27 $\pm$ 0.01 \\
            ARF & \cellcolor{green!30}1.98 $\pm$ 0.96 & \cellcolor{green!30}2.18 $\pm$ 0.54 & \cellcolor{green!30}3.87 $\pm$ 0.54 & \cellcolor{gray!30}0.96 $\pm$ 0.01 & \cellcolor{green!30}0.97 $\pm$ 0.004 & \cellcolor{green!30}0.99 $\pm$ 0.003 & 0.13 $\pm$ 0.004 \\
            Vines & \cellcolor{green!30}2.10 $\pm$ 0.87 & \cellcolor{green!30}5.18 $\pm$ 0.94 & \cellcolor{gray!30}4.80 $\pm$ 0.55 & \cellcolor{green!30}0.98 $\pm$ 0.008 & \cellcolor{green!30}0.97 $\pm$ 0.005 & \cellcolor{green!30}0.99 $\pm$ 0.004 & 0.14 $\pm$ 0.004 \\
            \bottomrule
        \end{tabular}
        }
    %}
    \vspace{1em}
    \caption{Results on Wisconsin breast cancer data. For each evaluation metric, the down (up) arrow means the lower (higher) the better respectively. All results are reported as the mean \(\pm\) std, with $x \pm y$ format.}
    \label{tab:wisconsin_comparison}
\end{table*}

%We generated $100$ datasets containing $1000$ samples for each method.
The results in Table \ref{tab:wisconsin_comparison} indicate that every generator has sufficient privacy guarantees i.e the DCR larger than $10 \%$ and that GMCM, Vines and ARF generators are outperforming the others generators in terms of \textit{fidelity} to the original dataset. Notably, the GMCM excels in metrics such as MMD and IKL, while also being simpler and more interpretable compared to a random forest-based model, and to vines copulas.
\subsection{Cleveland Heart Data}
\label{sec:cleveland}

The Cleveland Clinic Heart Disease Dataset is part of the UCI Machine Learning Repository. It contains \(303\) instances with $14$ attributes, including age, sex, blood pressure, cholesterol levels, and others related to heart health. The target attribute indicates the presence or absence of heart disease. We selected \(4\) variables: trestbps, thalach, chol and age.  

It is important to note that copula-based methods typically perform best with a limited number of variables, though we could have chosen more than four variables in this example.

\begin{table*}[h]
    \renewcommand{\arraystretch}{2.5}
    \resizebox{1.0\linewidth}{!}{
        %\Large
        %\scalebox{1.8}{
        \begin{tabular}{lccccccc}
        \toprule
        Generator Type & MMD (e-3) $\downarrow$ & MMDa (e-3) $\downarrow$  & JSD (e-3) $\downarrow$ & IKL (e-1) $\uparrow$  & P-PR  $\uparrow$  & P-RE  $\uparrow$  & DCR (e-2)  $\uparrow$  \\
        \midrule
        GC      & \cellcolor{gray!30}4.3 $\pm$ 0.0008  & \cellcolor{red!30}11.7 $\pm$ 1.2  & \cellcolor{red!30}9.7 $\pm$ 0.58  & \cellcolor{red!30}8.3 $\pm$ 0.52   & \cellcolor{red!30}0.98 $\pm$ 0.003  & \cellcolor{red!30}0.94 $\pm$ 0.008  & 0.34 $\pm$ 0.009 \\
        CTGAN   & \cellcolor{gray!30}4.30 $\pm$ 0.0009 & \cellcolor{red!30}54.96 $\pm$ 2.05  & \cellcolor{red!30}22.62 $\pm$ 0.60  & \cellcolor{red!30}8.26 $\pm$ 0.64   & \cellcolor{red!30}0.822 $\pm$ 0.01           & \cellcolor{red!30} 0.99 $\pm$ 0.003  & 0.53 $\pm$ 0.01\\
        GMCM    & \cellcolor{gray!30}4.30 $\pm$ 0.00002& \cellcolor{green!30}4.69 $\pm$ 0.0005 & \cellcolor{green!30}5.24 $\pm$ 0.0004 & \cellcolor{green!30}9.30 $\pm$ 0.01   & \cellcolor{green!30}0.98 $\pm$ 0.003           & \cellcolor{green!30}0.97 $\pm$ 0.01  & 0.32 $\pm$ 0.009 \\
        MARG    & \cellcolor{gray!30}4.30 $\pm$ 0.00001 & \cellcolor{red!30}10.01 $\pm$ 0.0009  & \cellcolor{green!30}5.35 $\pm$ 0.0004  & \cellcolor{green!30}9.31 $\pm$ 0.01   & \cellcolor{gray!30}0.96 $\pm$ 0.0006          & \cellcolor{gray!30}0.98 $\pm$ 0.009  & 0.35 $\pm$ 0.009\\
        ARF     & \cellcolor{gray!30} 4.30 $\pm$ 0.00002 & \cellcolor{green!30}2.35 $\pm$ 0.0004 & \cellcolor{green!30}4.45 $\pm$ 0.0005 & \cellcolor{green!30}9.76 $\pm$ 0.009 & \cellcolor{green!30}0.98 $\pm$ 0.005 & \cellcolor{green!30}0.99 $\pm$ 0.005 & 0.31 $\pm$ 0.008 \\
        Vines   & \cellcolor{gray!30}4.30 $\pm$ 0.00001 & \cellcolor{green!30}4.78 $\pm$ 0.0005  & \cellcolor{green!30}5.30 $\pm$ 0.0004  & \cellcolor{green!30}9.32 $\pm$ 0.01  & \cellcolor{green!30}0.98 $\pm$ 0.003           & \cellcolor{green!30}0.98 $\pm$ 0.01  & 0.33 $\pm$ 0.009 \\
         \bottomrule
        \end{tabular}
        }
    %}
    \vspace{1em}
    \caption{Results on Cleveland Heart dataset. For each evaluation metric, the down (up) arrow means the lower (higher) the better respectively. All results are reported as the mean \(\pm\) std, with $x \pm y$ format.}
    \label{tab:cleveland_comparison}
\end{table*}
The results presented in Table \ref{tab:cleveland_comparison} show that every generators has a DCR larger than $10 \%$. Furthermore, we can see that ARF, Vines and the GMCM performs well in terms of most of the metrics. %However, apart from this, the GMCM performs among the best generators in several metrics.

\section{Discussion}
\label{sec:discussion_synthetic}
In this paper, we presented a versatile statistical framework for generating tabular synthetic data based on copulas, which enable modelling of multivariate distributions with diverse marginals — a common scenario in medical data. To the best of our knowledge, this is a first instance of employing the Gaussian Mixture Copula Model (GMCM) for synthetic data generation, extending the copula-based literature to a model capable of capturing multi-modal dependency structures.

We applied this framework to a concrete use case: enabling probabilistic prediction of cardiovascular risk when some inputs to the risk calculator are systematically missing. To guide evaluation, we established specific \textit{desiderata} for synthetic data tailored to this problem and assessed quality along the three essential dimensions of \textit{fidelity}, \textit{utility}, and \textit{privacy}. Our framework yielded a sound empirical balance across all three dimensions, illustrating the value of a use-inspired perspective when assessing synthetic data generation methods.

The present results suggest that copula-based methods offer a practical compromise between utility and privacy, particularly well-suited to low-dimensional datasets with limited training observations. Unlike more complex generators such as GANs \cite{zhao2021ctab}, VAEs \cite{ishfaq2018tvae}, or diffusion models \cite{kotelnikov2023tabddpm}, which risk overfitting and potential privacy breaches in simpler settings, copula-based approaches remain robust in these contexts — as illustrated by the difficulties CTGAN encounters with small sample sizes, see Section \ref{sec::application} and \ref{sec:synthetic_experiments}. This work does not aim to provide a universal synthetic data generation method but instead advocates for the use of simpler statistical generators in low-dimensional tabular data scenarios.

Furthermore, copula-based methods for synthetic data generation have practical advantages. Compared to black-box approaches such as the Adversarial Random Forest (ARF), they offer greater interpretability: practitioners can incorporate domain expertise by specifying, for instance, the expected number of mixture components, and can inspect fitted marginals and dependence parameters directly. More broadly, copulas provide a full probabilistic model of the data, unlocking capabilities that go beyond synthetic data generation itself — including direct computation of quantities of interest, conditional sampling \cite{faul2024easy}, and uncertainty quantification.

Moreover, this work introduces new evaluation metrics for privacy assessment. Specifically, we adapted an attribute inference attack (AIA) \cite{jia2018attriguard} metric based on nearest neighbours to evaluate vulnerability to sensitive attributes, extending previous work to continuous attributes \cite{patki2016synthetic}. Notably, we applied this metric to assess an attribute not directly observed in the data but derived from a combination of other attributes — specifically, the SCORE2 risk \cite{esc2021score2}.

To assess the risk of replicating real patients in synthetic datasets, we used the DCR metric, a widely used tool in synthetic data generation \cite{lu2019empirical} literature but which may fail to detect privacy leakage \cite{yao2025dcr}. Additionally, while DCR provides insights into privacy, it lacks specific guidelines on when a dataset can be deemed sufficiently private.

To address this, we developed a permutation test that compares the closest distances within real datasets to those between real and synthetic datasets. The premise is that if both datasets originate from the same distribution, these distance vectors should share similar distributions. The approach is flexible in the choice of distance metric and test statistic definition; in this work, we opt for the Gower distance and use the difference in the minima of the two vectors as the test statistic. Additional details on the construction of this permutation test can be found in Appendix~\ref{sec::permutation}.

Applying this test to the case study presented in Section~\ref{sec:case_study}, we find that copula-based synthetic datasets yield large p-values, indicating that the null hypothesis is not rejected and suggesting adequate privacy protection.

\begin{figure}
    \centering
    \includegraphics[width = 1.02\columnwidth]{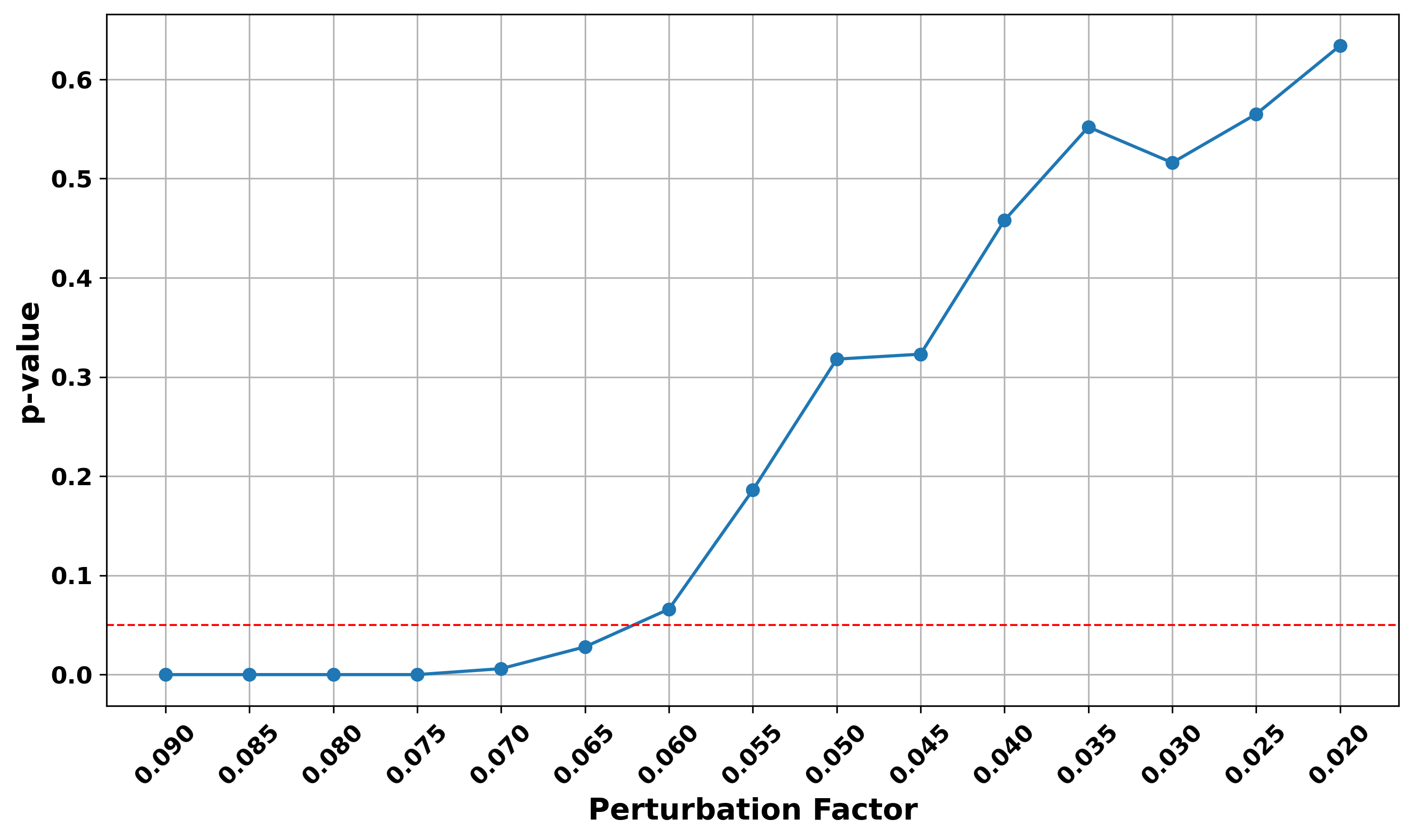}
    \caption{p-values of our permutation test after perturbing marginally the original dataset proportionally to the standard deviation by using different factors. }
    \label{fig:permutation}
\end{figure}

To further illustrate the behaviour of the test, we consider synthetic datasets obtained by adding varying levels of noise to each marginal of the real data from Section~\ref{sec:case_study}. As shown in Figure~\ref{fig:permutation}, noise levels below $0.1$ times the standard deviation result in p-values under $0.05$. This indicates, with $95\%$ confidence, that the closest distances between real and synthetic data points are smaller than those within the real dataset alone — a pattern that signals a potential privacy risk.

Note that this test is in its preliminary stages and is subject to further refinement. Our test is using the notion of identifiability introduced in  Yoon and al. \cite{yoon2020anonymization}, which measures if generated synthetic patient observations are “different
enough” from real patients by using distances between real patients as baseline.

Furthermore, in this work, we only considered similarity-based privacy metrics (SBPM) while some works have shown that they are inadequate to measure the resilience against privacy attacks and provide no formal guarantees against information leakage \cite{ganev2025inadequacy}. They advocate for the use of membership inference attacks (MIA) \cite{shokri2017membership} to measure privacy but their examples are over-simplistic and do not apply to our application.

A limitation of our case study is that the participants are from a relatively homogeneous population of women aged 40–69 years. It should also be noted that the present work evaluates the reconstruction of SCORE2 estimates rather than actual cardiovascular outcomes. Future studies should investigate whether synthetic-data-based approaches preserve predictive performance for hard clinical endpoints such as myocardial infarction, stroke or cardiovascular mortality.

While we advocate for the use of copula-based synthetic data generators in situations where practitioners handle small, low-dimensional datasets that cannot be shared due to privacy concerns, we also caution against the excessive reliance on synthetic datasets. The rapid advancements in diffusion models and transformers have enabled to perform data augmentation easily, allowing for the training of large machine learning models on synthetic data. However, the quality and privacy of these datasets are often inadequately assessed, leading to an influx of information online where the authenticity of data can be unclear.
In conclusion, the optimal approach is to use real data whenever feasible and adapt statistical analyses accordingly. Synthetic datasets should be utilized and published with caution.

Finally, it is important to highlight that current approaches for managing categorical variables and missing values are relatively simplistic, presenting avenues for future improvement. Specifically, a large number of categories or multiple categorical variables can result in complicating copula fitting (see Appendix \ref{sec::discrete}). Therefore, we recommend avoiding the discretization of continuous variables whenever possible, despite the common practice in medical applications of categorizing continuous measurements to distinguish pathological from non-pathological states.

Moreover, this work does not address other motivations for generating synthetic data, such as sampling from underrepresented classes in imbalanced datasets or creating fair synthetic data from biased sources \cite{van2021decaf}. Furthermore, beyond statistical similarity, future work should incorporate clinical plausibility constraints. In medical applications, synthetic patients should not only reproduce statistical properties but also respect biologically plausible combinations of clinical variables.
\newpage
\bibliographystyle{abbrv}
\bibliography{references.bib}

\appendix

\section{Other generators}
\label{sec::other_generators}
\subsection{ARF}
The Adversarial Random Forest (ARF) \cite{watson2023adversarial} is an algorithm that leverages random forests to estimate a multivariate distribution and generate synthetic data. The core idea is to partition the data space such that within each leaf, the data is approximately jointly independent.

This approach allows ARFs to require significantly fewer computational resources than other generative models, as the multivariate data density can be estimated through separate estimations of the marginal densities for each leaf. This simplifies the task of learning a joint distribution, especially as dimensionality increases. However, ARF models may involve deeper trees, although this typically does not present issues in practice \cite{watson2023adversarial}.

The ARF algorithm begins by fitting an unsupervised random forest (URF) \cite{shi2006unsupervised} to the data, creating an initial synthetic dataset $\Tilde{X}^{0}$ consisting of $n$ observations, sampled independently from the real data marginals $\mathbf{X}$. We then calculate the coverage of each leaf relative to the original data, defined as the ratio $\frac{n_{\ell}^{b}}{n_{b}}$ of data points in a leaf $\ell$ from tree $b$. A new synthetic dataset $\Tilde{X}^{1}$ is generated by sampling from marginals within random leaves, selected with probability proportional to this coverage. A random forest classifier $f^{(1)}$ is then fitted to distinguish $\mathbf{X}$ from $\Tilde{X}^{1}$. If this classifier's accuracy is sufficiently low, convergence is achieved; otherwise, the process iterates until convergence is reached.

For a learned ARF $f$, for each tree $b$, we record the split criteria $\theta_{b}$ and empirical
coverage $q(\theta^{\ell}_{b})$ of each leaf. We estimate distribution parameters $\psi^{\ell}_{b,j}$
independently for each variable $X_j$ in leaf $\ell$.
The estimated density is given by:

$$q(\mathbf{x}) = \frac{1}{B} \sum_{\ell, b : \mathbf{x} \in \mathcal{X}^\ell_b} q(\theta^\ell_b) \prod_{j=1}^{d} q(x_j; \psi^\ell_{b,j})$$

To generate synthetic data from this density, one samples a tree uniformly from $[B]$, selects a leaf based on $q(\theta^{\ell}{b})$, and samples each feature $X_j$ independently using the density parameterized by $\psi^{\ell}_{b,j}$.

The ARF is particularly suited for mixed tabular data settings and involves only a few hyperparameters.

\subsection{CTGAN}

A Generative Adversarial Network (GAN) typically comprises two neural networks: a generator \( G \) and a discriminator \( D \). The generator, \( G(z; \theta_g) : \mathcal{Z} \to \mathcal{X} \), takes random noise \( z \) from a latent space \( \mathcal{Z} \) and generates synthetic data that approximates the real data distribution in space \( \mathcal{X} \). Generally, both \( \mathcal{Z} \) and \( \mathcal{X} \) exist within \(\mathbb{R}^d\)and the noise \( z \) is often sampled from a standard normal distribution, denoted as \( p_Z(z) \).

The discriminator, \( D(x; \theta_d) : \mathcal{X} \to [0,1] \), receives either real data or data from \( G \) and outputs the probability of the data being real. By training \( G \) and \( D \) adversarially, the models aim to reach a minimax equilibrium described by:

\[
\min_{\theta_g} \max_{\theta_d} V(G, D) = \mathbb{E}_{x \sim p_X}[\log D(x; \theta_d)] + \mathbb{E}_{z \sim p_Z}[\log(1 - D(G(z; \theta_g)))]
\]

The conditional tabular generative adversarial network (CTGAN), introduced by Xu and Veeramachaneni in 2019 \cite{xu2019modeling}, addresses the challenges of generating tabular data with GANs. This architecture enhances GANs with mode-specific normalization to handle multi-modal continuous distributions. CTGAN also features a conditional generator that facilitates training-by-sampling. Data is generated by conditioning on a discrete column and sampled based on the discrete variable's observed categories. In sum, CTGAN's approach allows it to generate balanced datasets and handle categorical data more effectively by focusing on specific conditions when generating new samples.

As CTGANs are tailored for mixed tabular data types, they are regarded as a state-of-the-art method in synthetic data generation \cite{kiran2024methodology}. Their implementation is accessible through the Synthetic Data Vault library in \textit{Python}  \cite{patki2016synthetic}.

\section{Permutation Tests}
\label{sec::permutation}

Consider two samples: a real dataset of size \( n \), denoted by \(\mathbf{x}^{(\mathbf{r})} = (x^{(r)}_{1}, \ldots, x^{(r)}_{n})\), and a synthetic dataset of size \( n-1 \), denoted by \(\mathbf{x}^{(\mathbf{s})} = (x^{(s)}_{1}, \ldots, x^{(s)}_{n-1})\). For \( i \in \{1, \ldots, n\} \), define the intra-sample distance as:

\[
d^{(\text{intra})}_{i} = \min_{\substack{j \in \{1, \ldots, n\} \\ i \neq j}} d(x^{(r)}_{i}, x^{(r)}_{j})
\]

and the inter-sample distance as:

\[
d^{(\text{inter})}_{i} = \min_{j \in \{1, \ldots, n-1\}} d(x^{(r)}_{i}, x^{(s)}_{j})
\]

where \( d \) is the Gower distance.

Consider the vector of intra-sample minimum distances:

\[
\mathbf{d}_{\text{intra}} = \begin{pmatrix}
d^{(\text{intra})}_{1} \\
\vdots \\
d^{(\text{intra})}_{n}
\end{pmatrix}
\]

and the vector of inter-sample minimum distances:

\[
\mathbf{d}_{\text{inter}} = \begin{pmatrix}
d^{(\text{inter})}_{1} \\
\vdots \\
d^{(\text{inter})}_{n}
\end{pmatrix}
\]

If \(\mathbf{x}^{(\mathbf{r})}\) and \(\mathbf{x}^{(\mathbf{s})}\) are i.i.d. samples from the same distribution, \(\mathbf{d}_{\text{intra}}\) and \(\mathbf{d}_{\text{inter}}\) should have the same distribution. However, if the generator overfits the data, \(\mathbf{d}_{\text{inter}}\) will have smaller values, because points are copied in the synthetic dataset.

The test statistic is defined as:

\[
T = \min_{j \in \{1, \ldots, n\}} d^{(\text{intra})}_{j} - \min_{j \in \{1, \ldots, n\}} d^{(\text{inter})}_{j}
\]

A large \( T \) suggests that the smallest distance between a real and a synthetic point is less than between real points, leading us to reject the null hypothesis that both datasets originate from the same distribution.

Alternatives, like the difference in means between \(\mathbf{d}_{\text{intra}}\) and \(\mathbf{d}_{\text{inter}}\) can be considered but are less effective at detecting single copies in the synthetic dataset.

To perform the permutation test, we combined \(\mathbf{d}_{\text{intra}}\) and \(\mathbf{d}_{\text{inter}}\) into a single vector, shuffled it, and split it back into two vectors of size $n$ to calculate \( T \). Repeating this process \( N = 10,000 \) times provides the distribution of \( T \) under the null hypothesis. The rank of the observed test statistic within this distribution gives the \( p \)-value.

\section{Handling Discrete Variables}
\label{sec:discrete}

In copula-based synthetic data generation, handling discrete variables involves transforming them into continuous variables, applying the copula methods, and subsequently converting them back into discrete values. This approach introduces randomness to the transformation and is similar to the methodology outlined in \cite{benali2021mtcopula}, detailed as follows:

\begin{itemize}
    \item \textbf{Case 1: Ordinal Categorical Variables}
    \begin{enumerate}
        \item Estimate the proportions of each level (e.g., k levels) in the original dataset.
        \item Identify quantiles $z_1, \ldots, z_{(k-1)}$ of the standard normal distribution that partition the area under the normal curve to match these probabilities. Thus, $(-\infty, z_1]$ matches the first level, $(z_1, z_2]$ matches the second level, and so on, with $(z_{(k-1)}, \infty)$ matching level k.
        \item For each observation at level j, replace it with a random value from its truncated standard normal distribution with support $(z_{(j-1)}, z_j]$ (where $z_0$ is $-\infty$ and $z_k$ is $+\infty$).
        \item Convert all ordinal variables to continuous variables using this method.
        \item Apply the copula-based method to generate a synthetic dataset.
        \item Convert the generated variables back to ordinal categorical variables by reversing the transformation.
    \end{enumerate}
    \item \textbf{Case 2: Unordered Categorical Variables}
    \begin{enumerate}
        \item Convert the categorical variable into $k-1$ dichotomous variables using dummy coding. For example, $D_1$ is 1 if the level is 2 and 0 otherwise, $D_2$ is 1 if level is 3 and 0 otherwise, $\ldots$, $D_{(k-1)}$ is 1 if in level k and 0 otherwise (level 1 serves as the reference category and is coded as 0 across all dummies).
        \item Transform all dummy variables $D_1, D_2, \ldots, D_{(k-1)}$ into continuous variables as described in steps 2-4 above.
        \item Use the copula-based method to generate a synthetic dataset.
        \item Revert the generated synthetic variables $D_1, D_2, \ldots, D_{(k-1)}$ back to an unordered categorical variable by reversing the process in step 1.
    \end{enumerate}
\end{itemize}

It is important to note that for unordered categorical variables, a large number of categories or multiple categorical variables can lead to an explosion in the number of dummy variables. This makes fitting a copula in high dimensions challenging, which is a drawback of the method. However, in our experiments, we primarily dealt with ordinal categorical variables.
\label{sec::discrete}
\section{Marginal evaluation}
\label{sec::marginal_eval}

In this section, we analyze the marginal distributions of the synthetic datasets generated from the CIMBOLIC data (see Section \ref{sec::application}) using metrics such as mean and standard deviation. %To determine whether the marginal distributions significantly differ between real and synthetic datasets, we utilize statistical tests: the Student’s t-test for continuous variables and the Pearson Chi-square test for binary variables.

Table \ref{tab:marginals} presents the findings. Notably, only the CTGAN method produces synthetic datasets with marginal distributions that significantly deviate from those of the real datasets. All other generators appear to accurately capture the marginal distributions, which is anticipated for copula-based methods (GMCM, GC, and Vines), as well as the MARG method, due to their use of kernel density estimation for estimating marginals.
\begin{table*}[ht]
    \centering
    \renewcommand{\arraystretch}{2.5}
    \resizebox{1.05\textwidth}{!}{%
    \begin{tabular}{l|c||c|c|c|c|c|c}
        \toprule
        \textbf{Feature Name} & \textbf{REAL} & \textbf{ARF } & \textbf{GMCM $^{*}$} & \textbf{GC} & \textbf{MARG $^{*}$} & \textbf{VINES $^{*}$} & \textbf{CTGAN} \\
        \midrule
        Age &  50.87 $\pm$ 6.05 & 50.85$\pm$  6.01 & 50.79 $\pm$ 6.092 & 50.83 $\pm$ 5.91& 50.79 $\pm$ 6.07 & 50.79 $\pm$ 6.07  & 45.97 $\pm$  5.85 $^{*}$\\
        \hline
        BMI & 25.50 $\pm$ 5.25 &  25.43 $\pm$  5.26 & 25.35 $\pm$ 5.21 & 25.39 $\pm$ 4.93 & 25.33 $\pm$ 5.20 & 25.36 $\pm$ 5.21   & 23.83 $\pm$ 7.03  \\
        \hline
        BP & 124.19 $\pm$  16.68 & 123.48$\pm$ 17.2 & 123.82 $\pm$ 16.45 & 126.18 $\pm$ 8.40 &  123.9 $\pm$ 16.39 & 123.83 $\pm$ 16.38& 110.61 $\pm$ 24.40 \\
        \hline
        TotalC & 5.39 $\pm$ 0.98 & 5.40 $\pm$  0.99 & 5.37 $\pm$ 0.99 & 5.60 $\pm$ 0.28  & 5.37  $\pm$ 0.98 & 5.37 $\pm$  0.98& 4.93 $\pm$ 1.37 $^{*}$   \\
        \hline
        HDLC &   1.80 $\pm$ 0.58 &  1.83 $\pm$ 0.57& 1.78 $\pm$ 0.56 & 1.78 $\pm$ 0.54 & 1.78 $\pm$ 0.57 & 1.78 $\pm$ 0.56  & 1.27 $\pm$ 0.65\\
        \hline
        Smoker & 20.7\% & 21.2\% & 20.79\% &  21.8\% & 21.2\% & 20.5\% & 27.9\%   \\
        \hline
        Diabetes & 8.7\% & 10.3\% & 9.0\% & 9.5\% & 8.8\% & 8.8\% & 25.3\%  \\
        \hline
        SCORE2 & 2.14 $\pm$ 1.62 & 2.17 $\pm$ 1.74& 2.16 $\pm$ 1.64& 2.16 $\pm$ 1.39 & 2.22 $\pm$ 1.51 &2.16 $\pm$ 1.62 & 2.22 $\pm$  1.99 \\
        \bottomrule
    \end{tabular}%
    }
    \vspace{2em}
    \caption{Comparison of the feature distributions between the original dataset and synthetic datasets (generated by various methods). For continuous variables, we report the mean $\pm$ std and for binary variables, we report the proportion with the characteristic. $^{*}$ these $3$ generators use the same method to fit the marginal distributions.}%$^{*}$ denotes that the generated samples are statistically different from the original samples with significance level $\alpha=0.05$. }
    \label{tab:marginals}
\end{table*}

\section{Fitting a GMCM}
\label{sec::fitting_gmcm}
In this section, we will particularly focus on estimation of parameters for a Gaussian Mixture Copula model (GMCM).
Estimating parameters of copulas and in particular of the GMCM is a challenging task for several reasons as it has been discussed in \cite{tewari2023estimation, kasa2022improved, bilgrau2016gmcm}.

The first challenge related to copulas is that they are invariant under marginal increasing transformations.%, see Proposition \ref{prop::copula_invariance}. %Specifically, if $\mathbf{X} = (X_{1},\ldots,X_{d})^{t}$ is a random vector with continuous marginal distributions and copula $C$, and $T_{1},\ldots,T_{d}$ are strictly increasing functions. Then $(T_{1}(X_{1}),\ldots,T_{d}(X_{d}))^{t}$ also has copula $C$.

In the case of a GMCM, this property implies a lack of identifiability of the parameters, which makes the fitting of the parameters more difficult. This can be seen through the following proposition \cite{tewari2023estimation}:

\begin{proposition}(Non identifiability of GMCM)

    Let $\theta = \{ \alpha_{i},\mu_{i}, \Sigma_{i}\}_{i=1}^{K}$ and $\theta'= \{ \alpha_{i},A \mu_{i} + \mathbf{b}, A^{t} \Sigma_{i} A\}_{i=1}^{K}$ two valid set of parameters with $\mathbf{b} \in \mathbb{R}^{d}, ~A \in \mathbb{R}^{d \times d} $. Then the likelihood of the two Gaussian mixture copula models with these parameters are the same i.e $\forall \mathbf{u} \in \mathbb{R}^{d}, ~c_{gmc}(u_{1},\ldots,u_{d};\theta) = c_{gmc}(u_{1},\ldots,u_{d};\theta')$
\end{proposition}

This lack of identifiability results in the likelihood of a GMCM having infinitely many maximizers. To address this issue,  Bilgrau \cite{bilgrau2016gmcm} proposed standardizing the first component to have zero mean and unit variance along each dimension, which resolves the identifiability issue in most cases but some problems of identifiability
persist in a number of scenarios. Tewari \cite{tewari2023estimation} shows that identifiability is ensured under some constraints and proposes to use a suitable prior during parameter estimation to satisfy these constraints.

Another challenge in fitting GMCM is the high computational cost associated with functional evaluations, as the quantile function of a GMM isn't known in closed form \cite{tewari2011parametric}. To address this, we can use numerical approximation schemes leveraging the monotonicity of the cdf, although they tend to be computationally expensive. Alternatively, methods based on the bisection algorithm, such as the one developed by Chandrupatla \cite{chandrupatla1997new}, can be employed to find the inverse of a cdf and show improved computational performances. Finally, an other approach developed in \cite{kasa2020gaussian} consists in approximating this quantile function by using a first-order Taylor expansion of the Gaussian cdf.

In the literature, two primary methods have been proposed for estimating the parameters of a GMCM via direct gradient-based optimization. The classical approach involves using finite difference methods to estimate the derivatives. This approach approximates the derivatives numerically, which can be computationally expensive and less precise. Recently, a more efficient method, leveraging modern probabilistic programming language tools such as automatic differentiation, has been developed \cite{tewari2023estimation}. Automatic differentiation computes the exact values of the derivatives by leveraging the chain rule, significantly enhancing both computational efficiency and accuracy.

Another possible approach for parameter estimation is using an Expectation-Maximization (EM) type algorithm. However, due to the changing inverse distribution functions along the margins with every parameter update, the inputs of the EM algorithm are not fixed, rendering the traditional EM algorithm unsuitable. Although pseudo-EM algorithms have been proposed and discussed in the literature \cite{tewari2011parametric,bhattacharya2014unsupervised}, they optimize a pseudo-likelihood and often lack convergence guarantees, leading to suboptimal performance. These algorithms require additional checks at each iteration to ensure a monotonically increasing likelihood function. Some solutions have been suggested \cite{bhattacharya2014unsupervised,kasa2020gaussian}, primarily focusing on preventing abnormal drops in likelihood between EM steps.

In high-dimensional settings, imposing constraints on covariance matrices using regularized Parsimonious Gaussian Mixture Models (PGMM) \cite{mcnicholas2008parsimonious}, combined with an Alternating Expectation Conditional Maximization (AECM), appears to yield promising results \cite{kasa2020gaussian}. Recently, it has been demonstrated \cite{tewari2023estimation} that, through some detailed but straightforward manipulation, the expectation of the complete data log-likelihood can be expressed in closed form. However, the maximization step cannot be performed analytically, necessitating gradient-based optimization and resulting in a Generalized EM algorithm.

Most of these methods have been implemented in \textit{R} inside the package \textit{GMCM} \cite{bilgrau2016gmcm}. Meanwhile, the automatic differentiation method has been developed in \textit{Python} by Tewari.

\section{Synthetic experiments}
\label{sec:synthetic_experiments}
We generated data from a $2$-dimensional GMCM with random parameters. The marginal distributions have been chosen to be respectively Gaussian and Gamma distributed.

\begin{table*}[h]
    \renewcommand{\arraystretch}{2.5}
    \resizebox{1.0\linewidth}{!}{
        \Large
        %\scalebox{1.8}{
        \begin{tabular}{lccccccc}
            \toprule
            Generator Type & MMD (e-3) $\downarrow$ & MMDa (e-3) $\downarrow$ & JSD (e-3) $\downarrow$ & IKL $\uparrow$ & P-PR $\uparrow$ & P-RE $\uparrow$ & DCR $\uparrow$\\
            \midrule
            GC & \cellcolor{red!30}37.30 $\pm$ 3.190 & \cellcolor{red!30}22.29 $\pm$ 2.28 & \cellcolor{red!30}3.84 $\pm$ 0.07 & \cellcolor{black!10}0.98 $\pm$ 0.01 & \cellcolor{red!30}0.76$\pm$ 0.01 & \cellcolor{red!30}0.99 $\pm$ 0.003 & 0.19 $\pm$ 0.019 \\
            CTGAN & \cellcolor{red!30}50.06 $\pm$ 2.77 & \cellcolor{red!30}39.13 $\pm$ 2.77 & \cellcolor{red!30}11.02 $\pm$ 0.77 & \cellcolor{red!30}0.94 $\pm$ 0.01 & \cellcolor{red!30}0.74 $\pm$ 0.01 & \cellcolor{red!30}0.99 $\pm$ 0.002 & 0.25 $\pm$ 0.018 \\
            GMCM & \cellcolor{green!30}3.54 $\pm$ 0.77 & \cellcolor{green!30}2.12 $\pm$ 0.66 & \cellcolor{black!10}3.71 $\pm$ 0.68 & \cellcolor{black!10}0.99$\pm$ 0.005 & \cellcolor{green!30}0.95 $\pm$ 0.005 & \cellcolor{green!30}0.99 $\pm$ 0.005 & 0.08 $\pm$ 0.004 \\
            MARG & \cellcolor{red!30}51.54 $\pm$ 3.30 & \cellcolor{red!30}34.62 $\pm$ 2.44 & \cellcolor{black!10}3.61 $\pm$ 0.74 & \cellcolor{black!10}0.99$\pm$ 0.006 & \cellcolor{red!30}0.70 $\pm$ 0.01 & \cellcolor{red!30}0.99 $\pm$ 0.001 & 0.26 $\pm$ 0.025 \\
            ARF & \cellcolor{green!30}0.86 $\pm$ 0.34 & \cellcolor{green!30}0.68 $\pm$ 0.35 & \cellcolor{green!30}2.89 $\pm$ 0.61 & \cellcolor{black!10}0.99 $\pm$ 0.004 & \cellcolor{green!30}0.98 $\pm$ 0.004 & \cellcolor{green!30}0.99 $\pm$ 0.001 & 0.06 $\pm$ 0.004\\
            Vines & \cellcolor{red!30}37.03 $\pm$ 2.79 & \cellcolor{red!30}21.85 $\pm$ 2.02 & \cellcolor{black!10}3.58 $\pm$ 0.58 & \cellcolor{black!10}0.99 $\pm$ 0.006 & \cellcolor{red!30}0.75 $\pm$ 0.01 & \cellcolor{red!30}0.99 $\pm$ 0.002 & 0.20 $\pm$ 0.01 \\
            \bottomrule
        \end{tabular}
        }
    %}
    \vspace{1em}
    \caption{Experiment Results. For each evaluation metric, the down (up) arrow means the lower (higher) the better respectively. All results are reported as the mean \(\pm\) std, with $x \pm y$ format.}
    \label{tab:simulated_comparison}
\end{table*}

Table \ref{tab:simulated_comparison} indicates that GMCM and ARF are the top-performing models for generating synthetic data according to the selected metrics. However, they both have a DCR score lower than $10 \%$ indicating potential privacy concerns. Thus for this example none of the synthetic generators can be recommended for use.

\section{Experimental details}
\label{sec:experimental_details_synthetic}
\textbf{Vines.}
After transforming the data to uniform marginals non-parametrically using kernel density estimation with a Gaussian kernel, 
we utilised the \textit{R} package \textit{VineCopula} to select a vine structure by considering  R-vines with possible families Gaussian or Student. We then simulate $1000$ points from the fitted vine copula model.

\vspace{1em}

\textbf{Gaussian Copula synthetizer.}
We used the function \textit{GaussianCopulaSynthesizer} of the \textit{Python} package \textit{SDV} with default arguments 
 to fit and sample $1000$ synthetic data points.

\vspace{1em}

\textbf{CTGAN.} We used the function \textit{CTGANSynthesizer} of the \textit{Python} package \textit{SDV} with $500$ epochs  to fit and sample $1000$ synthetic data points.

\vspace{1em}

\textbf{ARF}  We used the function \textit{arf} of the \textit{Python} package \textit{arfpy} with default arguments to fit the adversarial random forest to our datasets. 

\vspace{1em}

\textbf{MARG} We used the function \textit{densityfun} from the \textit{R} package \textit{statip} to perform kernel density estimation, employing a Gaussian kernel with default settings. This process was repeated for each marginal, followed by sampling from each.

\end{document}